\documentclass[article,a4paper,twocolumn,showpacs,superscriptaddress,amssymb]{revtex4}
\usepackage{bm}
\usepackage{amssymb}
\usepackage{amsfonts}
\usepackage{amsmath}
\usepackage{graphicx}
\usepackage[dvips]{color} 

\definecolor{g-blue}{rgb}{0.83,0.95,1}
\definecolor{Blue}{rgb}{0.5,0.5,1}
\definecolor{DarkBlue}{rgb}{0.00,0.00,0.58}
\definecolor{g-yellow}{rgb}{1,1,0.7}
\definecolor{g-green}{rgb}{0.9,1,0.9}
\definecolor{green}{rgb}{0,0.6,0}
\definecolor{Green}{rgb}{0,0.4,0}
\definecolor{cyan}{rgb}{0,0.7,0.7}
\definecolor{black}{rgb}{0,0,0}
\definecolor{grey}{rgb}{0.4 ,0.4 ,0.4 }

\def\be{\begin{equation}}\def\ee{\end{equation}}
\def\bea{\begin{eqnarray}}\def\eea{\end{eqnarray}}
\def\bse{\begin{subequations}}\def\ese{\end{subequations}}
\newcommand{\BE}[1]{\begin{equation}\label{#1}}
\newcommand{\BEA}[1]{\begin{eqnarray}\label{#1}}
\newcommand{\BSE}[1]{\begin{subequations}\label{#1}}

\let \= \equiv \let\*\cdot \let\~\widetilde \let\^\widehat \let\-\overline

  \def\1{\bm1} 

\def\<{\left\langle}    \def\>{\right\rangle}
\def\({\left(}          \def\){\right)}
 \def \[ {\left [} \def \] {\right ]}

 \renewcommand{\O}{\Omega}

\newcommand{\vf}{V_{\mathrm f}}

\def\Oms{\Omega_{\rm s}}
\def\Vf{V_{\rm f}}

\def\aen{\alpha_{\rm en}}
\def\taam{\tilde\alpha_{\rm am}}
\def\taen{\tilde\alpha_{\rm en}}

\begin{document}

\title{Kelvin-Helmholtz instability of AB interface in superfluid $^3$He}

\author{V.B.~Eltsov$^1$, A.~Gordeev$^1$, M.~Krusius}

\address{Department of Applied Physics, Aalto University, P.O. Box 15100, FI-00076 AALTO, Finland}

\date{\today}

\begin{abstract}

The Kelvin-Helmholtz instability is well-known in classical hydrodynamics, where it explains the sudden emergence of interfacial surface waves as a function of the flow velocity parallel to the interface. It can be carried over to the inviscid two-fluid dynamics of superfluids, to describe the stability of the phase boundary separating two bulk phases of superfluid $^3$He in rotating flow, when the boundary is localized with a magnetic field gradient. The results from extensive measurements as a function of temperature and pressure confirm that in the superfluid the classic condition for stability is changed and that the magnetic polarization of the B-phase at the phase boundary has to be taken into account, which yields the magnetic field dependent interfacial surface tension.



\vspace{3mm}



\end{abstract}
\maketitle

\section{Introduction}
\label{Intro}

The Kelvin-Helmholtz instability (KH) is one of the celebrated instabilities of classical hydrodynamics \cite{HydroDynamics}. The traditional example is that of two horizontal fluid layers of different densities flowing at different velocities parallel to their common interface. This state of relative laminar shear flow is stable at low velocities, \textit{i.e.} the interface remains flat and smooth, but an instability in the form of an interfacial surface wave develops when the difference in the velocities reaches a critical value. Originally \cite{LordKelvin,Helmholtz} the two flows were assumed inviscid, but countless manifestations of the KH instability in different systems illustrate that the instability also survives in viscous settings.

In superfluids the phase boundary separating the two bulk-liquid phases of superfluid $^3$He, the A and B phases, provides an extraordinary opportunity for examining interfacial dynamics. Here the flow of the superfluid components can be truly inviscid, the difference in mass density is negligible, as the interface is formed as a sharp but continuous change in the order parameter of the superfluid state, with a width on the order of the superfluid coherence length of (10---100) nm, depending on the liquid pressure. Measurements have shown \cite{KH-Instability} that the KH instability does not lead to an oscillating response, but owing to large damping from orbital viscosity, only to a bulge in the interface contour which protrudes into the B phase and enables the formation of quantized vortices in the B phase. The newly created vortices reduce the flow velocity to subcritical levels at the interface and the bulge decays. This phenomenon has been explained by reformulating the classical instability condition for the superfluid case \cite{Volovik}, by taking into account the coupling to the fixed reference frame via the normal component. Here we summarize the results over a broad range of measurements and examine their agreement with the superfluid instability condition.

\section{KH instability in superfluid $^3$He}
\label{KH_in_3He}

\begin{figure}[t!]
\centerline{\includegraphics[width=1.1\linewidth]{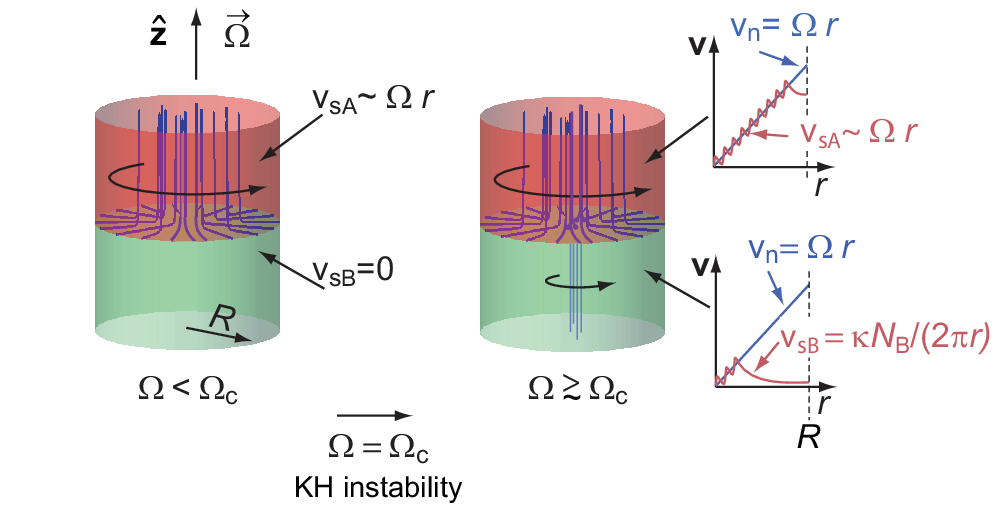}}
\caption{KH instability in superfluid $^3$He. The two bulk phases of superfluid $^3$He can coexist in a cylinder rotating at constant angular velocity $\Omega$ in an inhomogeneous axially oriented "barrier`` magnetic field $H_\mathrm{b}(z) \boldmath{\hat {z}}$. The interface is at the location $z_0$, where $H_{\mathrm b} (z_0) = H_\mathrm{AB} (T,P)$: in the high-field region $H_\mathrm{b} \geq H_\mathrm{AB}$ A phase is stable (top section) while at lower fields $H_\mathrm{b} \leq H_\mathrm{AB}$ B phase is stable (bottom section). \textbf{\textsl{(Left)}} When $\Omega$ is increased at constant temperature from zero to some value less than $\Omega_\mathrm{c} (T,P)$, the A-phase section is filled with rectilinear double-quantum vortices at the equilibrium density $\Omega/\kappa$.  At the AB interface these curve to form a vortex sheet covering the interface, leaving the B-phase section vortex free. \textbf{\textsl{(Right)}} When $\Omega$ is increased past $\Omega_\mathrm{c}$, the KH instability mediates the transfer of a bundle of small vortex loops through the AB interface into the B-phase. In the temperature regime of laminar flow above $0.6\,T_\mathrm{c}$ these grow to rectilinear single-quantum vortices which extend  across the AB interface and form a cluster at the equilibrium density $2 \Omega/\kappa$ in the centre of the cylinder (as shown here). In the turbulent regime below $0.6\,T_\mathrm{c}$ the loops interact in a turbulent burst, filling the B-phase section ultimately with the equilibrium number of rectilinear vortices. }
\label{Principle}
\end{figure}

 Our KH measuring arrangement is illustrated in Fig.~\ref{Principle}. The two-phase configuration is stabilized in an inhomogeneous magnetic field which exceeds the critical field $H_\mathrm{AB} (T,P)$ above which $^3$He-A becomes energetically preferable to $^3$He-B at temperature $T$ and pressure $P$. The initial state in rotation is engineered to have the equilibrium number of quantized vortices in the A phase, providing its solid-body-like rotation, while the superfluid component in the B phase is vortex free and thus non-rotating (stationary in the laboratory frame). This meta-stable non-dissipative state persists to relatively high rotation, until at a critical rotation velocity $\Omega_\mathrm{c}$  an interfacial mode is excited and a number of vortices from the A phase manage to cross into the B phase. The sudden appearance of new B-phase vortices is the experimental signal for the instability. The dependence of $\Omega_\mathrm{c}$ on rotation, temperature, pressure, and magnetic field gradient at the AB interface is measured in these experiments.

The meta-stable starting situation is reached by increasing rotation slowly at constant temperature from rest to some angular velocity $\Omega$ which is below $\Omega_\mathrm{c}$. The A-phase section of the long sample cylinder has by then become filled with rectilinear doubly-quantized vortices \cite{DoubleQuantumVortex}. These vortices have an extended ``soft" core (of radius $\sim 10\,\mu$m),  with a continuous order parameter distribution in a skyrmion texture, and correspondingly a low critical rotation velocity for their formation. At the AB interface their double-quantum cores dissociate, bend parallel to the interface, and extend radially out to the cylinder wall, forming thereby a vortex sheet which covers the interface \cite{InterfaceVortexSheet}. In contrast, B phase vortices have an order of magnitude higher critical velocity and are not formed below $\Omega_\mathrm{c}$ of the KH instability. The reason \cite{KH-Review} is the narrow ``hard" core of the B phase vortex with a radius comparable to the superfluid coherence length $\xi_0 \approx \hbar v_\mathrm{F} / (k_\mathrm{B} T_\mathrm{c}) \gtrsim 10\,$nm (where $v_\mathrm{F}$ is the Fermi velocity and $T_\mathrm{c}$ the superfluid transition temperature). Thus the B phase section on the left in Fig.~\ref{Principle} remains vortex free. This meta-stable starting configuration persists because energetically a sizeable local concentration of kinetic energy would be required to constrict the two orders of magnitude fatter core of the A-phase vortex to a narrow B phase core.

When the rotation drive $\Omega$ is next increased by a small increment $\Delta \Omega$ to $\Omega_\mathrm{c}$ (where $\Delta \Omega \ll \Omega$) or slightly above, the phase boundary loses stability and a surface wave is triggered. It unleashes the escape of a variable number of small vortex loops, each carrying a single quantum of superfluid circulation $\kappa = h/(2m_3)$. These loops protrude from the A phase vortex sheet into the B phase and lie initially closely packed, covering one surface depression of the interfacial wave \cite{KH-Review}. At high temperatures $T > 0.6\,T_\mathrm{c}$ the large mutual friction damping in vortex motion limits interactions among the loops and their number remains constant while they grow to rectilinear vortex lines. In the final state they are arranged as a central cluster of rectilinear singly-quantized B-phase vortices, as seen on the right in Fig.~\ref{Principle},  with a topological ``boojum" point defect at the AB interface where they connect to dissociated halves of the A-phase vortex \cite{PointDefect}.

Towards lower temperatures the mutual friction damping $\alpha (T,P)$ decreases \cite{Makinen}, ultimately exponentially in the B phase. Below $0.6\,T_\mathrm{c}$, $\alpha \lesssim 1$ and the B phase enters the turbulent regime. Here the number of vortex loops escaping within the interfacial surface depression is not conserved while the loops evolve to rectilinear B-phase vortex lines. Instead the closely-packed loops interact and proliferate via reconnections to a turbulent burst \cite{TurbulentTransition}. In the rotating counterflow the turbulent burst ultimately evolves \cite{Front} to the equilibrium vortex state, where the number of vortex lines $N_\mathrm{eq} \approx \pi R^2 \, n_\mathrm{v}$. Here $R$ is the radius of the cylinder and $n_\mathrm{v} = 2\Omega/\kappa$ the areal density of rectilinear lines in the B-phase vortex array in solid-body rotation.

 Experimentally it is the number of vortex lines which is monitored continuously non-invasively with NMR spectrometers. Their detector coils are located outside the rotating sample cylinder above and below the AB interface, displaced far enough from the interface so that they reside in homogeneous axially oriented polarizing magnetic fields.

In helium liquids the superfluid KH instability was first invoked as an explanation for the threshold to capillary wave formation on the free surface of superfluid $^4$He \cite{Korshunov}. A comparison to measurements was later performed in Ref.~\cite{Remizov}. The KH instability has also been proposed as a tool to explore the interface of the superfluid dilute $^3$He-$^4$He liquid mixture with its normal concentrated phase \cite{Burmistrov}. A flurry of theoretical investigations have appeared which recommend the KH instability as a means to study mixtures of different cold-atom Bose-Einstein condensates (see eg. \cite{Kobyakov,Lundh,Suzuki,Ohta}). Another intriguing suggestion is to use the KH instability for modelling a hydrodynamic analogue of the event horizon of a gravitational black-hole \cite{Volovik2,VolovikBook}.

In neutron star physics the superfluid KH instability has emerged as a possible mechanism by which equilibrium can be reached in one  ``glitch" between the angular momentum reservoirs of the superfluid and normal fractions of a spinning neutron star \cite{Melatos}. The sudden discontinuous glitch-like ``spin-up event" in the neutron star's rotation velocity has been compared to the angular velocity jump which one expects in a KH instability event when the outward directed motion of quantized vortices across an interface between two different neutron condensates is calculated. Estimates of the superfluid fraction and its degeneracy temperature in the interior of a neutron star indicate that superfluidity extends across the boundary of the inner crust and the central core \cite{Andersson}. Unfortunately, the phase diagram of neutron superfluids in a strong density gradient much above nuclear densities is largely unknown. The spherical geometry of the neutron star's interior complicates the KH response from that of the simple example in Fig.~\ref{Principle}. This is illustrated by superfluid $^3$He measurements where the AB interface is a near concentric cylindrical surface within the rotating cylinder \cite{Krusius}.

Here we start with a brief overview of the superfluid KH model at the AB phase boundary, reminding about its analytical definition in Sec.~\ref{KH instability} and about experimental aspects in Sec.~\ref{Experimental}. For further details we refer to Ref.~\cite{KH-Review}. Measurements on the critical rotation velocity $\Omega_\mathrm{c}$ are summarized in Sec.~\ref{Results}. In Sec.~\ref{Analysis} the implications from the analysis are discussed. The results display good agreement with the theory for the KH instability and provide a powerful illustration of the $^3$He superfluids as quantum model systems.

\section{KH Theory}
\label{KH instability}

The classic example of the KH-instability takes place at the horizontal interface of two inmiscible and inviscid fluid layers with densities $\rho_1$ and $\rho_2$. Assigning $v_1$ and $v_2$ as the corresponding flow velocities parallel to the interface, the instability develops when the relative velocity $\mid v_1 - v_2 \mid$ satisfies Lord Kelvin's condition \cite{LordKelvin}
\begin{equation}
 {\rho_1 \rho_2 \over \rho_1+\rho_2} (v_1-v_2)^2=2\sqrt{\sigma F},
\label{InstabilityCriterion}
\end{equation}
where the interface is characterized by its surface tension $\sigma$ and its restoring force $F$ which often is the gravitational force $F = g (\rho_2 - \rho_1)$. The interface becomes unstable when the free energy of the perturbed interface drops below that of the flat surface and an interfacial capillary wave is formed with the wave vector
\begin{equation}
k_0=\sqrt{F/\sigma}~. \label{WaveVector}
\end{equation}

\subsection{Instability at low magnetic field}
\label{ZeroField}

At an interface separating two superfluids the situation is different since the normal component also becomes important. It has two roles: 1) as a source of dissipation in interface motion and 2) in establishing a link between superfluid flow and the external reference frame. The dissipation arises from quasiparticle scattering from the sharp order parameter anomaly at the interface \cite{Kopnin} and from orbital viscosity when the interface is set into motion \cite{Leggett}. The external reference frame is represented by the velocity $v_{\rm n} = \,\mid \mathbf{\Omega} \times \mathbf{r} \mid$ of the normal component which enters as a third velocity in the instability condition \cite{Volovik}. It couples the superfluid fractions to the reference frame of the cylinder rotating at constant angular velocity $\mathbf{\Omega}$, so that solid-body corotation with the container walls is preferred. In the Galilean invariant form the instability condition becomes
\begin{equation}
{1\over 2}\rho_{\rm s1}~ (v_{\rm s1}-v_{\rm n}) ^2 +{1\over
2}\rho_{\rm s2}~ (v_{\rm s2}-v_{\rm n}) ^2
    =  \sqrt{\sigma F}~.
\label{SuperfluidInstability}
\end{equation}
We see here that the superfluid instability becomes possible also when the relative superfluid velocity vanishes, $v_{\rm s1} = v_{\rm s2}$, and the two streams parallel to the interface flow at the same velocity. The shape of the AB interface is maintained by the interfacial surface tension $\sigma$ and the restoring force $F$ arising from the inhomogeneous magnetic stabilizing field: $F= {1 \over 2} \, (\chi_\mathrm{A} - \chi_\mathrm{B} ) \, \nabla H^2 \approx (\chi_\mathrm{n} - \chi_\mathrm{B}) \, H_z \, (dH_z/dz)$, where $\chi_\mathrm{n} \approx \chi_\mathrm{A}$ and $\chi_\mathrm{B}$ are the normal phase, A phase, and B phase susceptibilities. The deformation of the interface starts with the wave vector $k_0$ given by Eq.~(\ref{WaveVector}).

 A simplification can be introduced in Eq.~(\ref{SuperfluidInstability}) since the critical velocity of spontaneous vortex formation in the A phase is low, of order 0.1\,rad/s \cite{CritVelocityA-Phase}, while in the B phase it is much higher $\gtrsim 2\,$rad/s, depending on the smoothness of the cylinder wall \cite{VorForm}. We approximate the A phase as being in the equilibrium rotating state, $\mathbf{v}_\mathrm{sA} \approx \mathbf{\Omega} \times \mathbf{r} = \mathbf{v}_\mathrm{n}$ (in the laboratory coordinate system), so that only the flow velocity in the vortex-free B phase counts: $\mathbf{v}_\mathrm{sB} = - \mathbf{\Omega} \times \mathbf{r}$ (in the rotating coordinate system). Eq.~(\ref{SuperfluidInstability}) is thereby reduced to
\begin{equation}
\Omega_{\mathrm{c}}^2 =  2 \frac {\sqrt{\sigma F}} {\rho_{\rm sB} \; R^2}\,.
\label{SuperfluidInstability-2}
\end{equation}
Here $\rho_{\rm sB}$ is the density of the superfluid component in the B phase at the interface. Note that compared to the corresponding value from Eq.~(\ref{InstabilityCriterion}) for the ideal inviscid fluid, $\Omega_{\mathrm{c}}^2$ in Eq.~(\ref{SuperfluidInstability-2}) is smaller by a factor of 2.

For analysing measurements, the instability criterion can be divided in predominantly experimental and theoretical parts, $w_\mathrm{exp}$ and $w_\mathrm{theo}$,
\begin{equation}
w_\mathrm{exp} = \frac{(\Omega_{\mathrm{c}}\, R_\mathrm{eff})^4} {2\,\mid \nabla H^2 \mid_{H=H_\mathrm{AB}}} =  \frac {\sigma \; \Delta \chi} {\rho_\mathrm{sB}^2} = w_\mathrm{theo} \,.
\label{SuperfluidInstability-4}
\end{equation}
Here $R_{\rm eff} = v_\mathrm{sB}^\mathrm{c} / \Omega_\mathrm{c} $ is used for the radial location of the instability site at the interface. In Ref.~\cite{InterfaceVortexSheet} it was experimentally determined to be displaced from the cylinder wall, at  $R_{\rm eff} \approx 0.87 \, R$. The left side of Eq.~(\ref{SuperfluidInstability-4}) contains quantities which we determine experimentally, while on the right the three quantities $\sigma$, $\Delta \chi = \chi_\mathrm{A} - \chi_\mathrm{B}$, and $\rho_\mathrm{sB}$ we obtain from the literature or calculate from their theoretical expressions.

From Eqs.~(\ref{WaveVector}) and (\ref{SuperfluidInstability-2}) the wave length $\lambda$ of the over-damped interfacial excitation mode created at the instability is seen to be
\begin{equation}
\lambda = 4 \pi \; {\sigma \over \rho_\mathrm{sB}} \; (\Omega_\mathrm{c} \, R_\mathrm{eff})^{-2} \,.
\label{WaveLength}
\end{equation}
This quantity obtains experimental significance in the measurement of Fig.~\ref{Principle}. The instability is signaled by the transfer of vortices across the interface, when vortex loops escape from an interfacial surface wave depression which protrudes on the B-phase side \cite{KH-Review}. Initially the escaping vortices are part of the interfacial vortex sheet and coat the surface wave depression, \textit{i.e.} the number $\Delta N$ of such vortex loops is that which fits in one half of the wave length $\lambda$ of the surface excitation mode. The instability is a complex non-equilibrium event, but the escaping circulation $\Delta N$ is well-defined with a measurable statistical distribution function \cite{VorInjection}.

This can be confirmed at temperatures above the transition to turbulence, $T > 0.6\,T_\mathrm{c}$, where the number of loops expanding in the B-phase section to rectilinear vortex lines is conserved and thus can be counted in the final state after the instability. The number of circulation quanta, which in the interfacial vortex sheet flare out to the cylinder wall, is $N_\mathrm{eq} \approx \pi R^2 (2\Omega_\mathrm{c} /\kappa)$. Per unit length measured along the perimeter one then has $\sim R \, \Omega_\mathrm{c}/\kappa$ quanta flaring radially outward so that one half of the wave length of the surface wave fits $\Delta N \approx {1 \over 2} \lambda \, R_\mathrm{eff} \, \Omega_\mathrm{c}/\kappa$ quanta. Taking typical numbers, we arrive at an estimate $\lambda \sim 0.4\,$mm$^\dagger$. \footnotetext[1]{$^\dagger$For this estimate we use numbers pertinent for Fig.~\ref{KH-Steps}.}

Measurements on the instability condition~(\ref{SuperfluidInstability-2}) are conducted at the critical field $H_\mathrm{AB} (T,P)$. They are preferably performed at constant pressure, by either scanning temperature or the magnetic field gradient at constant temperature (\textit{i.e.} by changing $I_\mathrm{b}$, see Fig.~\ref{Setup}). At constant pressure $P$, $H_\mathrm{AB}$ increases with decreasing temperature, approaching values as large as $\lesssim 0.6\,$T at the lowest temperatures (the unit tesla (T) is used for magnetic field). With typical values for the superfluid $^3$He properties at low fields, reasonable agreement can be reached with Eq.~(\ref{SuperfluidInstability-2}) at high temperatures above about $T/T_\mathrm{c} > 0.7$---0.8, corresponding to $H_\mathrm{AB} \lesssim 0.1\,$T \cite{KH-Instability}. At lower temperatures the magnetic polarizability of the B phase reduces the stability of the AB interface and the measured $\Omega_\mathrm{c}$ falls increasingly below that calculated with low-field values (see illustration \textit{e.g.} in Fig.~14 of Ref.~\cite{KH-Review}). At $\sim 0.4\,T_\mathrm{c}$ the difference is of order $\sim 0.1\, \Omega_{\mathrm{c}}$.

\subsection{Magnetic field dependence}
\label{MagFieldDep}

Towards low temperatures $H_\mathrm{AB} (T,P)$ increases and introduces changes in the B-phase properties. To gain qualitative understanding, consider the Ginzburg-Landau (GL) expansion of the magnetic-field-induced gap deformation which in first order is parabolic $(\Delta_\parallel / \Delta_\mathrm{B0} )^2 \approx 1- 2\, a_1 H^2 \, (1-T/T_\mathrm{c}) $ and $(\Delta_\perp / \Delta_\mathrm{B0})^2 \approx 1 + {1 \over 4} a_1 H^2 \, (1-T/T_\mathrm{c})$. Here the anisotropic gap widths $\Delta_\parallel$ parallel and $\Delta_\perp$ perpendicular refer to the gap axis which lies along the unit vector $\hat{\bm \ell} = \hat{\mathbf H} \cdot \bar{\mathbf R}$,  where the unit vector $\hat{\mathbf H} = \mathbf{H}/H$ points in the direction of the applied field $\mathbf H$, and  $\bar{\mathbf R} (\hat{\mathbf n} , \theta) $ is a rotation matrix which rotates around the axis $\hat{\mathbf n}$ by the angle $\theta = \arccos {(- {1 \over 4}\Delta_\parallel / \Delta_\perp)}$. In the GL expansion of the gap the parabolic correction scales with pressure as $a_1 \propto T_\mathrm{c}^2$ \cite{Kleinert}. The corresponding suppression of the axial component of the superfluid density is  $\rho_{\mathrm{s}\parallel} / \rho_\mathrm{B0} =  1 - a_2 H^2 / (1-T/T_\mathrm{c}) + ... $ , where  $a_2 \propto a_1$.

Little quantitative experimental information exists on the field dependences of the relevant B-phase properties at temperatures below the GL regime, but the changes can be estimated with numerical weak coupling calculations \cite{Schopohl,Nagai}. At zero pressure in the zero-temperature limit such calculations give a gap distortion \cite{Nagai} which agrees well with vibrating wire measurements on the onset of pair-breaking as a function of the applied magnetic field, as reported by Shaun Fisher \textit{et al.} \cite{Fisher}. In comparison to the magnetic polarization effects, the depairing and gap suppression expected from superfluid flow is small in the range of the present KH measurements.

In addition, magnetic fields affect B-phase textural orientations and require a reconsideration of the boundary conditions at the AB interface. Generally in B-phase the order parameter texture has less of an influence on the hydrodynamic stability than in the A phase, where the texture at the wall of the rotating cylinder determines \textit{e.g.} the critical flow velocity of vortex formation \cite{CritVelocityA-Phase}. Nevertheless, orientational considerations lead to corrections which can be built into Eq.~(\ref{SuperfluidInstability-2}) in terms of a renormalized superfluid density $\rho_{\rm s,eff}$ \cite{AnisotropicKH}.

At the interface one requires 1) continuity of mass flow $\bm{\nabla} \cdot (\tensor{\rho}_\mathrm{sB} \cdot \mathbf{v}_\mathrm{sB} )= 0$ and 2) stability of the phase boundary such that no mass flow takes place through the interface in the direction of its unit normal $\hat{\mathbf s}$, \textit{i.e.} $\hat{\mathbf s} \cdot (\tensor{\rho}_\mathrm{sB} \cdot \mathbf{v}_\mathrm{sB}) = 0$. Writing the B phase mass density tensor in the form $\tensor {\rho}_\mathrm{sB} = \tensor {\rho}_\mathrm{s} = \rho_\mathrm{x} \; \hat{\mathbf{x}}\hat{\mathbf{x}} + \rho_\mathrm{y} \; \hat{\mathbf{y}}\hat{\mathbf{y}} + \rho_\mathrm{z} \; \hat{\mathbf{z}}\hat{\mathbf{z}} $ and choosing the cartesian coordinate $\hat{\mathbf{x}}$ to lie along the flow and along the interface while $\hat{\mathbf {z}}$ is the direction perpendicular to the interface, then Eq.~(\ref{SuperfluidInstability-2}) can be amended in the form
\begin{equation}
\Omega_{\mathrm{c}}^2 =  2 \frac {\sqrt{\sigma F}} {\rho_{\rm s,eff} \; R^2}\,,
\label{SuperfluidInstability-3}
\end{equation}
where the superfluid density $\rho_{\rm sB}$ has been replaced by an effective quantity \cite{AnisotropicKH},
\begin{equation}
\rho_{\rm s,eff} = \rho_{\rm x} \sqrt{\rho_\mathrm{x}/\rho_\mathrm{z}}\,.
\label{EffectiveDensity}
\end{equation}

The magnetic-field-distorted B-phase superfluid density is of the symmetric uniaxial form
\begin{equation}
\rho_{ij} = \rho_{\parallel}\; \hat{\mathbf {\ell}}_i \, \hat{\mathbf {\ell}}_j + \rho_{\perp}\; (\delta_{ij} - \hat{\mathbf {\ell}}_i \, \hat{\mathbf {\ell}}_j) ,
\label{DensityTensor}
\end{equation}
when expressed with respect to the anisotropy axis of the gap.


A number of interactions act to orient the anisotropy axis $\hat{\bm \ell}$, giving rise to an order parameter texture $\hat{\bm \ell}(\bm{r})$, which is of the axially symmetric `flare-out' configuration in a long cylinder with bulk B phase. At moderate magnetic fields in the regime of typical NMR measurements \cite{Textures}, the `flare-out' textures have been examined in numerous studies and the magnitude of the various textural interactions is well documented \cite{Thuneberg}. This applies to the B-phase textures within the two NMR detector coils in Fig.~\ref{Setup}, where the field is of order $H \sim 30\,$mT, \textit{i.e.} well above the field $H_{\mathrm D} \gtrsim 1$\,mT corresponding to the dipolar spin-orbit interaction, but well below the critical field $H_\mathrm{AB}$ needed for stabilizing the AB interface. The flare-out texture is formed by the coupling to the magnetic field, to rotation, and by the boundary condition at the cylinder wall. Owing to the presence of a gradient energy, in spatially inhomogeneous conditions these interactions have a characteristic range or `healing length'. These lengths have been measured in low-field NMR conditions, but at high fields of order of $H_{\mathrm AB}$, texture studies require extremely high field homogeneity and become experimentally demanding. It is not accurately known what is the relative range of the interactions determining the configuration of $\hat{\bm \ell}(\bm{r})$ in the vicinity of the AB interface and what alignment therefore should be assigned to $\rho_{\rm s,eff}$ in Eq.\,(\ref{EffectiveDensity}).

Close to the interface one finds regions in the B-phase texture where the orientation can be along any of the cartesian axes $x,y$, and $z$. As the dominant orienting interaction is the magnetic field, the gap axis is predominantly aligned parallel to the field, $\hat{\bm \ell} = \hat{\mathbf H} = \hat{\mathbf {z}}$, and the effective B-phase superfluid density in Eq.~(\ref{EffectiveDensity}) becomes
\begin{equation}
\rho_{ \hat{\bm \ell} \parallel \hat{\mathbf z}} = \rho_{\perp} \; \sqrt{\rho_{\perp}/\rho_{\parallel}}\;.
\label{EffectiveDensity-1}
\end{equation}
Since the isotropic zero magnetic field superfluid density $\rho_\mathrm{s0} \approx \rho_{\perp} > \rho_{\parallel}$, the effective superfluid density is magnified: $\rho_{\hat{\bm \ell} \parallel \hat{\mathbf z}} > \rho_\mathrm{s0} $. In the high-fields close to the interface the alignment $\hat{\bm \ell} \parallel \hat{\mathbf z}$ is expected to prevail and to occupy most of the flare-out texture. Right at the interface on the other hand, the boundary condition on $\hat{\bm \ell}$ requires parallel alignment to the AB interface \cite{InterfaceVortexSheet}. To minimize the flow energy, $\hat{\bm \ell}$ will then also align itself along the flow, $\hat{\bm \ell} = \hat{\mathbf {x}}$, and
\begin{equation}
\rho_{\hat{\bm \ell} \parallel \hat{\mathbf x}} = \rho_{\parallel} \; \sqrt{\rho_{\parallel}/\rho_{\perp}}\;.
\label{EffectiveDensity-2}
\end{equation}
 In this part of the texture, the effective superfluid density is reduced: $\rho_{\hat{\bm \ell} \parallel \hat{\mathbf x}} < \rho_\mathrm{s0} $. Alignment along the third cartesian axis, $\hat{\bm \ell} = \hat{\mathbf{y}} = \hat{\mathbf {r}}$, is realized in a boundary layer at the cylinder wall, enforced by a boundary condition which orients $\hat{\bm \ell}$ perpendicular to the wall, so that
\begin{equation}
\rho_{ \hat{\bm \ell} \parallel \hat{\mathbf r}} = \rho_{\perp} \;.
\label{EffectiveDensity-3}
\end{equation}
In this case the effective superfluid density remains practically unchanged from its low-field value since $ \rho_{\perp} \gtrsim \rho_\mathrm{s0} $.

Hence the B-phase order parameter texture may influence the hydrodynamic stability of the AB interface. The magnetic healing length $\xi_H (T,H)$, the length scale on which $\hat{\bm{\ell}}$ bends from $\hat{\bm \ell} = \hat{\mathbf H} = \hat{\mathbf {z}}$ in the bulk to $\hat{\bm \ell} = \hat{\mathbf {x}}$ at the interface, is obtained by comparing the magnetic orientation energy density of order $[ (\Delta_{\parallel} - \Delta_{\perp}) \Delta_{\perp} / H^2 ] \; (\hat{\bm{n}} \cdot \bm{H})^2$ to the gradient energy density, which resists distortions from uniform $\hat{\bm{\ell}}$ orientation.

At the low fields of conventional NMR measurements  $\xi_H $ is found to be $\propto 1/H$ and in magnitude a sizeable fraction of the sample cylinder radius $R$ \cite{Hakonen}. At higher fields accurate information on $\xi_H$ is lacking. Recent measurements on the dissipation recorded while the AB interface is oscillated with an ac magnetic field were explained assuming the heating to arise from orbital viscosity of the oscillating $\hat{\bm{\ell}}$ orientation \cite{Haley}. Good agreement with measured heating levels at different frequencies was obtained assuming a short healing length $\xi_H \sim 0.1\,$mm and a uniform bulk texture with $\hat{\bm \ell} = \hat{\mathbf H} = \hat{\mathbf {z}}$ in $H_\mathrm{AB} = 0.34\,$T magnetic field at zero pressure and $0.16\,T_\mathrm{c}$. This suggests that on moving away from the AB interface into bulk B phase the $\hat{\bm{\ell}}$ orientation recovers rapidly within a short distance $\ll R$ from being parallel to the interface to being oriented along the field. Owing to this parallel alignment $\hat{\bm \ell} = \hat{\mathbf {x}}$ at the interface, it is $\chi_{\perp} \approx \chi_\mathrm{B0}$ which enters in the magnetic restoring force $F$.

Actually, the development of the instability is a complex nonlinear phenomenon \cite{Lushnikov} which involves a range of length scales: $R \gg R - R_\mathrm{eff} \sim \lambda > \xi_H$. For instance, when the amplitudes of the perturbations sampling the interface stability become comparable to $\lambda$, they have already exceeded $\xi_H$ and the effective density increases to that in Eq.~(\ref{EffectiveDensity-1}). This reduces the critical velocity $\Omega_\mathrm{c}$. Experimentally, the site of the instability becomes the spot in the texture where the required rotation velocity is the lowest.

To explain the reduced interface stability at large fields, we note that at intermediate temperatures the enhancement $\rho_{\rm s,eff} / \rho_\mathrm{s0}$ amounts to several percent and increases towards low pressures, but at the lowest temperatures $\lesssim 0.2\,T_\mathrm{c}$ it vanishes exponentially. It therefore becomes evident that the changes in $\rho_\mathrm{sB}$ and $\chi_\mathrm{B}$ must be smaller compared to the reduction in the surface tension $\sigma$.

The interface tension is experimentally accessible only at the critical field $H_\mathrm{AB} (T,P)$. Being a sharp interface in the order parameter distribution with a width on the order of the superfluid coherence length $\xi (T,P) = \xi_0 / \sqrt{(1 - T/T_\mathrm{c})}$, the surface tension is of order $\sim \xi f_\mathrm{c}$. Here $f_\mathrm{c}$ is the condensation energy which in the GL regime can be expanded in the form $\propto (1 - T/T_\mathrm{c})^2 [1 - a_3 H^2 / (1- T/T_\mathrm{c})]$ \cite{Thuneberg_sigma}. Thus the surface tension is often expressed as $\sigma_\mathrm{AB} (T,P) \approx \sigma_0 (P)\, (1 - T/T_\mathrm{c})^{3/2}$ at pressures above the polycritical point where A phase is stable in zero field. This was demonstrated by Osheroff and Cross in their classic surface tension measurement at melting pressure \cite{Osheroff}. At high fields the surface tension $\sigma_\mathrm{AB} (T,H,P)$ has only been measured by Bartkowiak \textit{et al.} at zero pressure and $0.15\,T_\mathrm{c}$ \cite{Bartkowiak}. Its calculation is a more complex task, since one has to find the minimum-energy order parameter trajectory from A phase to the field-distorted B-phase energy minimum \cite{Kleinert_AB}. Overall, in increasing magnetic field the energy barrier is reduced and the surface tension decreases.

\vspace{5mm}
\section{Experimental method}
\label{Experimental}
\subsection{Measuring setup}

\begin{figure}[t]
  \centerline{
  \includegraphics[width=0.95\linewidth]{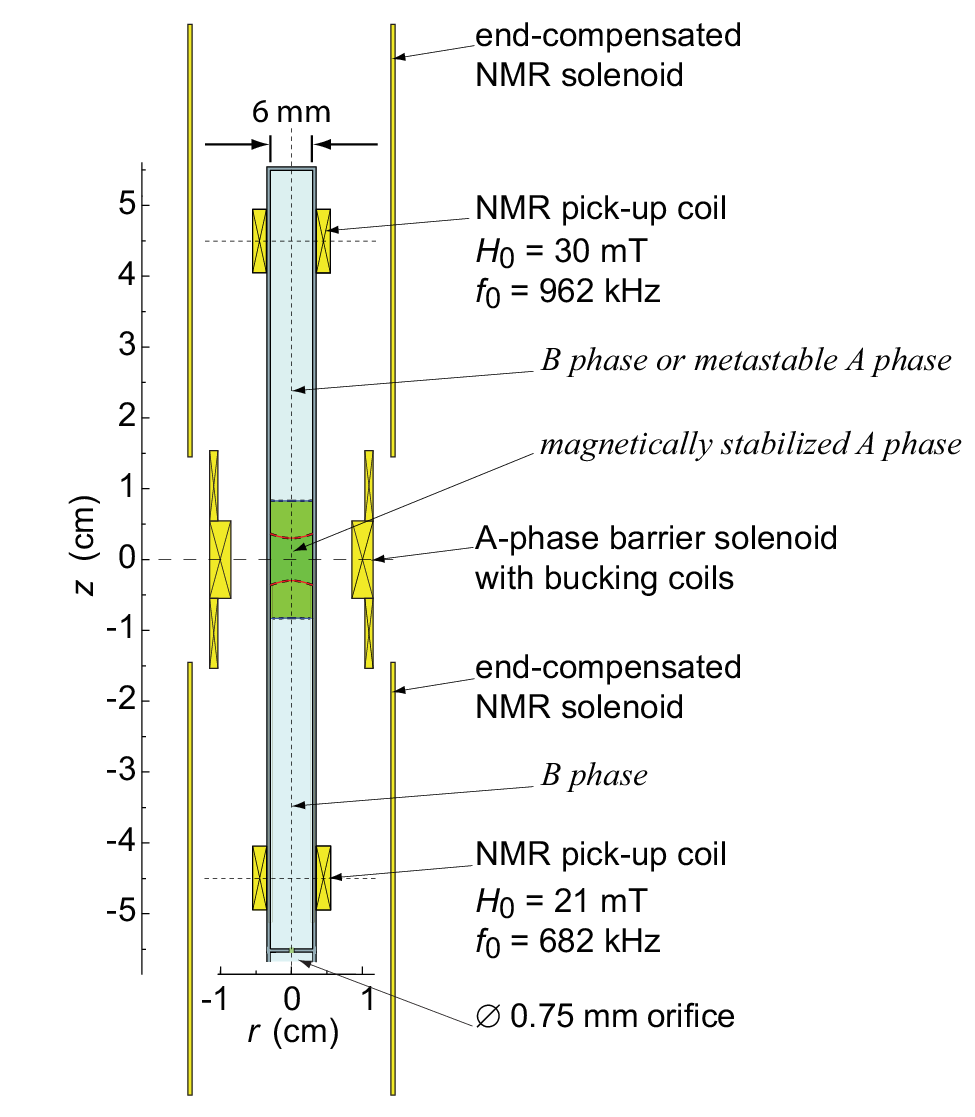}}
\caption{Measurement of KH instability. The liquid $^3$He sample is contained in a long quartz glass cylinder with smooth walls. The section in the middle is magnetically maintained in the A phase, while NMR detectors with single-vortex measuring resolution are located at both ends of the cylinder.  The axially oriented magnetic fields are generated with superconducting solenoids \cite{ExpSetup} which are thermally connected to the mixing chamber of the dilution refrigerator, used for precooling the copper nuclear cooling stage. A division plate with a small aperture separates the experimental volume from the thermal connection to the nuclear cooling stage, a sintered heat exchanger with rough surfaces.  Two examples of AB interfaces are shown at 29\,bar pressure and $0.55\,T_\mathrm{c}$: the inner concave phase boundaries are stabilized with a current $I_\mathrm{b} = 4\,$A in the barrier solenoid and the outer flatter boundaries with $I_\mathrm{b} = 8\,$A.    }
\label{Setup}
\vspace{-5mm}
\end{figure}

 Our KH measurements have been performed in the experimental setup of Fig.~\ref{Setup}. This is a versatile platform for different types of studies, if the placement of apertures and sensors is varied according to different needs. The heart of the setup is a fuzed quartz glass cylinder of 11\,cm length and 0.6\,cm inner diameter, which is used as sample container. A small superconducting solenoid around the middle section of the long tube carries a current $I_\textrm{b}$ and generates the magnetic field for stabilizing a layer of A phase which acts as a barrier between B-phase sections at each end of the tube. We call this the BAB stacking configuration of phases, in contrast to the metastable AB configuration where the entire top section above the lower AB interface is filled with A phase. Thus depending on magnetic field, temperature, pressure, and prehistory these two different configurations of A and B phases can be realized in the cylinder.

\begin{figure}[t]
  \centerline{
  \includegraphics[width=0.8\linewidth]{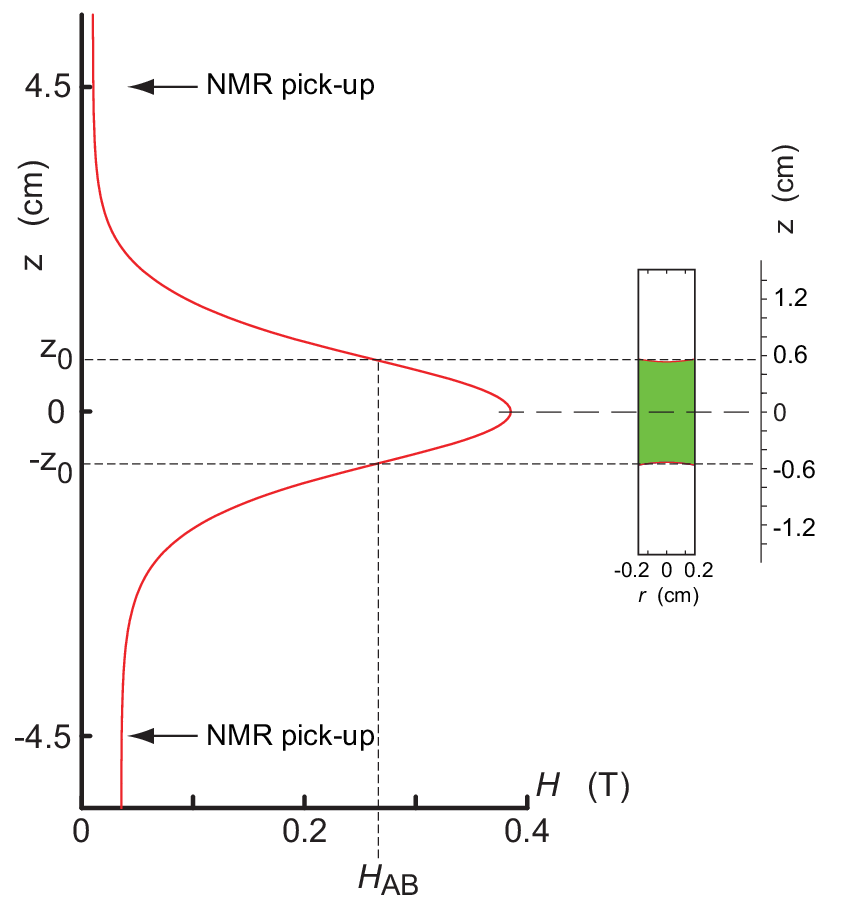}}
\caption{Axial distribution of the barrier magnetic field at $I_\mathrm{b} = 4\,$A \textbf{(on the left)}. \textbf{On the right} the A-phase layer in the BAB configuration is shown at 29\,bar pressure, $1.62\,$mK = $0.67\,T_\mathrm{c}$, and $H_\mathrm{AB} = 0.276\,$mT. The calculated shape of the concave AB interface meniscus \cite{LT23-FieldConf} is mainly determined by the magnetic energy (owing to $\Delta \chi = 5.25*10^{-8}\,$cgs) in the solenoidal field which increases radially, but also to a smaller extent by the surface tension $\sigma_\mathrm{AB} = 9.3\, \mu$erg/cm$^2$, by vortex-free rotation at 1\,rad/s in the B-phase sections and an equilibrium  vortex state in the A layer (with vortex formation starting at $\Omega_\mathrm{cA} = 0.15\,$rad/s). The contact angle at the cylinder wall is fixed to $\theta_\mathrm{AB} = 68^{\circ}$ .   }
\label{BarrierField}
\end{figure}

 Fig.~\ref{BarrierField} shows the axial distribution of the barrier field. The magnetic field is rapidly changing both within and outside the A-phase barrier layer, while further away at both ends of the sample tube end-compensated solenoids create homogeneous polarizing fields for monitoring the superfluid order parameter field with low-field NMR spectrometers. Low-field NMR is the realm  where experiment and theory of order parameter textures is well established \cite{Textures}. Here the relative amplitudes of the measured continuous-wave NMR signal provide information about the number of vortices and the frequency shifts of satellite peaks can be calibrated to provide a temperature reading.

In rotation the section with A phase is filled with approximately the equilibrium number of vortices in the form of lines or sheets with continuous order parameter distributions and low critical velocity ($\Omega_\mathrm{cA} \approx 0.1\,$rad/s) of vortex formation. In contrast, in the B-phase sections an important aspect is the surface quality of the inner cylinder wall, as the critical velocity of singular-core vortex formation depends crucially on the smoothness of the wall. With careful cleaning and etching, combined with visual screening in a microscope, isolated surface defects can be eliminated which makes critical velocities $\Omega_\mathrm{cB} \gtrsim 2\,$rad/s possible. In addition careful cool-down procedures are required, to avoid frozen water or gas accumulations on the walls. Moreover, to isolate from contact with the rough sintered heat exchanger surface below the sample tube, the cylinder is terminated with an orifice on the cylinder axis, which here has a diameter of 0.75\,mm. The upper limit of vortex-free rotation, which corresponds to an apparent effective velocity of vortex formation, was measured at 33.7\,bar pressure in the absence of the A-phase barrier layer (\textit{i.e.} when $I_\mathrm{b} = 0$). It proved to have a temperature independent value of $(2.2\,\pm 0.25)\,$rad/s in the range (0.55---0.75)$\,T_\mathrm{c}$. Thus a KH instability where $\Omega_\mathrm{c}$  exceeds this value, would not be accessible with this sample cylinder.

The obvious drawback from the orifice is the large thermal resistance which it presents to axial heat flow. Heat leaks of order 10 -- 100\,pW flowing from the sample cylinder through the orifice to the much colder heat exchanger volume (maintained at roughly $0.14\,T_\mathrm{c}$) limit the lowest achievable temperature to $\sim 0.20\,T_\mathrm{c}$ in the experimental volume above the orifice.

\begin{figure}[t]
  \centerline{
  \includegraphics[width=0.95\linewidth]{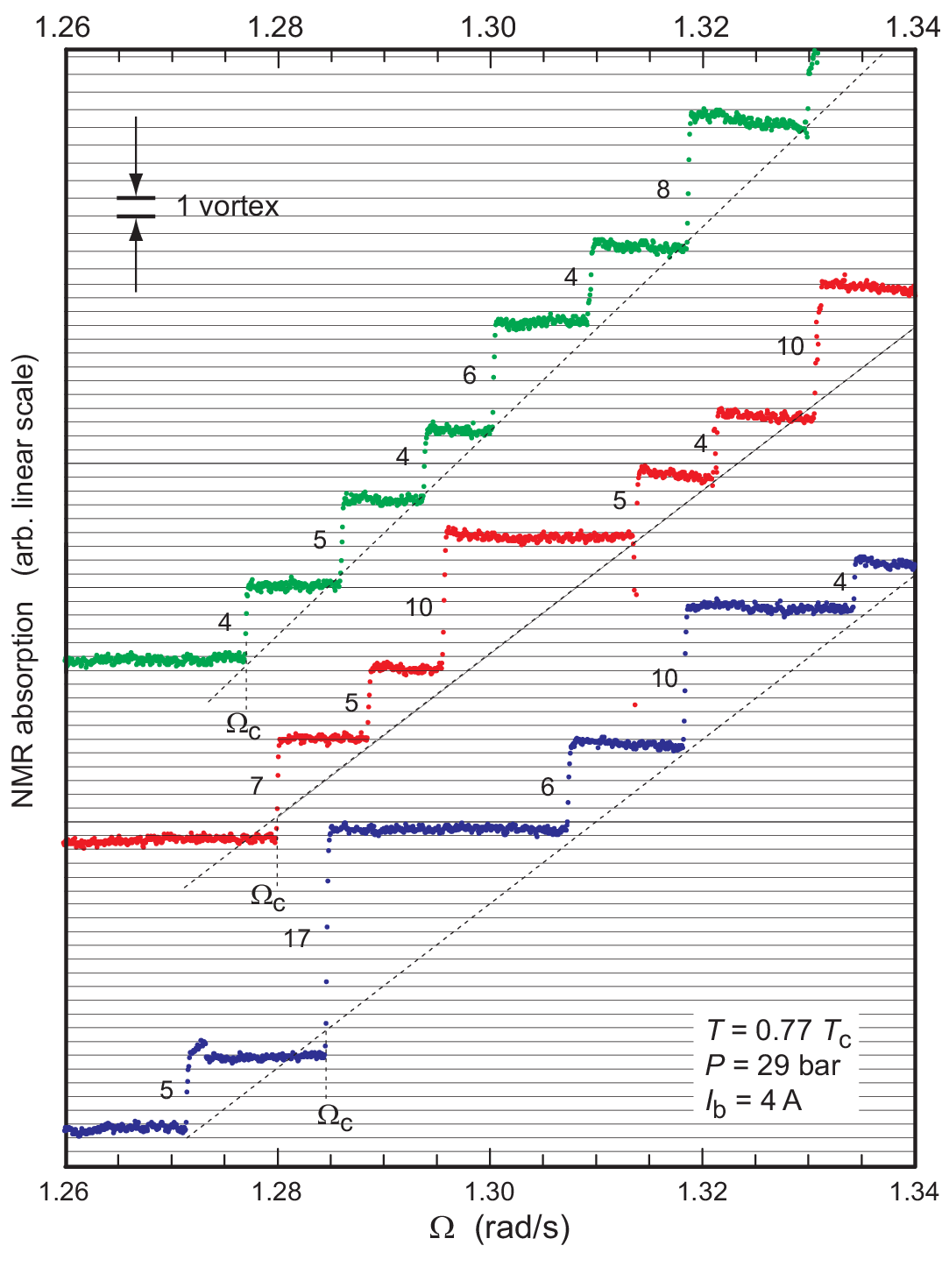}}
\caption{NMR signal from KH measurements. Three series of measurements are shown as a response to a slow increase of  rotation at a rate of $2*10^{-4}$\,rad/s$^2$. This is slow enough to maintain near equilibrium conditions. The NMR output plotted on the vertical axis monitors the absorption level at the Larmor edge of the NMR spectrum. In vortex-free rotation at $\Omega < \Omega_\mathrm{c}$ the NMR absorption is shifted in a counterflow peak, far from the Larmor edge. When $\Delta N$ vortices are created in a KH event, part of the absorption is shifted to the Larmor edge. At small vortex numbers $N \ll N_\mathrm{eq}$ the changes in absorption from the counterflow peak to the Larmor edge are approximately $\propto \Delta N$, as illustrated by the grid where the spacing corresponds to one vortex. The numbers next to each KH discontinuity give $\Delta N$.  }
\label{KH-Steps}
\end{figure}

\subsection{Measuring procedures}

We use continuous wave excitation for recording the NMR spectrum, by sweeping the polarization field. In a KH instability event the B-phase line shape changes discontinuously, as some NMR absorption is shifted to the Larmor edge, while the total integrated NMR absorption remains constant. As seen in Fig.~\ref{KH-Steps}, the resulting signal from a single instability event is quite prominent. Here the NMR absorption is recorded continuously close to the Larmor edge, where the absorption increases discontinuously when $\Omega$ is increased at a slow rate past the consecutive KH instabilities. The first discontinuity defines the critical velocity $\Omega_\mathrm{c}$, while the succession of the following new instability events defines the critical line
\begin{equation}
v_\mathrm{c} = \Omega_\mathrm{c} \, R_\mathrm{eff} = \left( \Omega - \frac {\kappa \, N(\Omega)} {2\pi R_\mathrm{eff}^2} \right) \, R_\mathrm{eff} \,,
\label{CritLine}
\end{equation}
where the KH critical flow velocity $v_\mathrm{c}$ is a constant and $N(\Omega)$ is the number of vortices which have broken through the AB interface by the time rotation has been increased to the value $\Omega \geq \Omega_\mathrm{c}$. The dashed line has been fitted to the critical end points of the staircase pattern and provides the slope $\Delta N/\Delta \Omega = 2 \pi R_\mathrm{eff}^2/\kappa$ which yields $R_\mathrm{eff} = \sqrt {{\kappa \over {2 \pi}} {{\Delta N} \over {\Delta \Omega} } } \approx 2.5\,$mm. It should be compared to the cylinder radius $R = 3\,$mm.

As seen in Fig.~\ref{KH-Steps}, the number of circulation quanta $\Delta N$ can be even or odd. Thus, despite the fact that the A phase is filled with doubly-quantized vortex lines, the circulation covers the AB interface on the A-phase side as a vortex sheet which is made up of single-quantum structures \cite{InterfaceVortexSheet}. On an average one finds that $\Delta N \approx 8$ \cite{InterfaceVortexSheet}. In the BAB stacking configuration, the instabilities of the two AB interfaces occur independently and randomly, but follow the same critical line. The scatter of the critical points in Fig.~\ref{KH-Steps} is related to the stability of the measuring conditions, \textit{i.e.} temperature and pressure, while inherent fluctuations do not appear to be influential. Thus the precision in Fig.~\ref{KH-Steps} can be improved by recording the full NMR spectra at constant rotation just before and after triggering an instability event with an incremental rotation increase by $\Delta \Omega$ which makes it possible to correct for drifts afterwards.

\begin{figure}[t]
  \centerline{
  \includegraphics[width=0.95\linewidth]{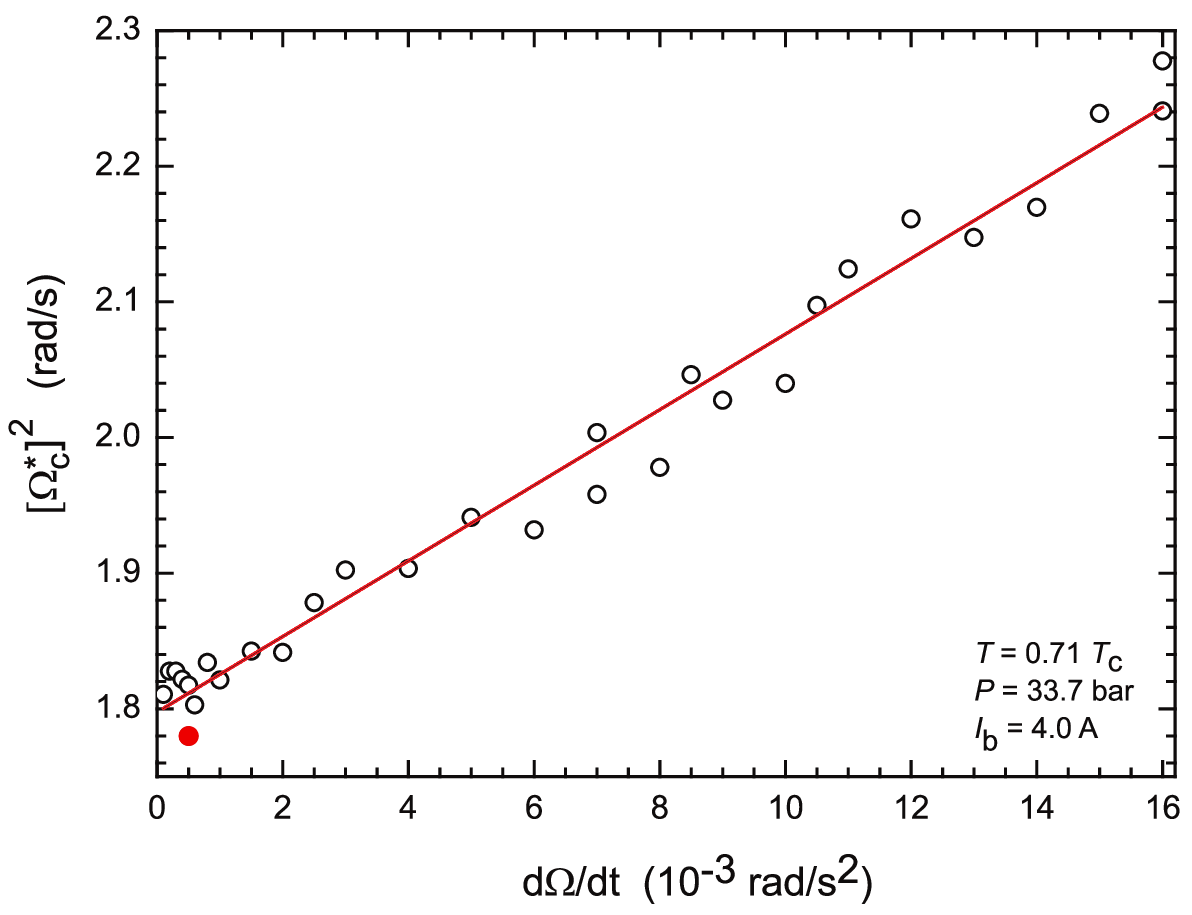}}
\caption{Response time of KH measurement. For each data point rotation is increased from zero at constant rate $\dot{\Omega}$, plotting the square of the apparent KH critical velocity $\Omega_\mathrm{c}^*$ as a function of the rate $\dot{\Omega}$, while $T= 0.71\,T_\mathrm{c}$, $P=33.7\,$bar, and $I_{\mathrm b}=4\,$A are maintained constant. The vertical intercept gives the true critical velocity $\Omega_\mathrm{c}$ and the slope the dissipative vortex mutual friction coefficient $\alpha$. The lowest data point (filled circle) 
was recorded while the A-phase vortex structure was the vortex sheet \cite{VortexSheet}. In all other cases the A-phase vortex structure was the doubly quantized vortex line \cite{DoubleQuantumVortex}. The response is independent of the A-phase vortex texture.    }
\label{AccelerationDependence}
\end{figure}

The response time of the measurement is limited by the velocity with which the information about an instability event travels to the NMR spectrometers, \textit{i.e.} by the expansion of the newly formed B-phase vortices from the AB interface to the NMR region. A single vortex line expands in vortex-free rotation such that it extends from the AB interface as a rectilinear line along the cylinder axis to its curved end which connects to the cylinder wall. The curved end precesses in spiral motion, moving away from the AB interface towards increasing counterflow.  The velocity of the spiral motion is largely determined by that section of the vortex end which is perpendicular to the cylinder wall \cite{FlightTime}. Its velocity has the axial and azimuthal components
\begin{equation}
\bm{v}_\mathrm{L} = - \alpha \, \Omega \, R \, \hat{\bm{z}} - (1-\alpha^{\prime})\, \Omega \, R \, \hat{\bm{\phi}} \,,
\label{SpiralMotion}
\end{equation}
where $\alpha (T)$ and $\alpha^{\prime}(T)$ are the dissipative and reactive mutual friction coefficients. Thus the axial expansion time is $\Delta t \approx d/(\alpha \Omega R)$ for bridging the distance $d$ from the AB interface to the NMR coil. This estimate is a good approximation even in the case when many vortices are expanding simultaneously. At high temperatures, for instance at $0.85\,T_\mathrm{c}$ at 29\,bar pressure where the AB interface is in thermodynamic equilibrium, the delay $\Delta t$ is a few seconds, but below $0.2 \, T_\mathrm{c}$ in the regime of ballistic quasiparticle motion where $\alpha (T)$ tends exponentially towards zero and the vortex response is turbulent, $\Delta t$ is more than an hour \cite{Makinen}.

In a measurement where rotation is increased at a constant rate $d\Omega /dt = \dot{\Omega}$, the apparent measured critical velocity $\Omega_\mathrm{c}^*$ is shifted higher by the equivalent of the delay,
\begin{equation}
[\Omega_\mathrm{c}^*]^2 = \Omega_\mathrm{c}^2 + \frac {2 d}{\alpha \, R} \dot{\Omega} \,.
\label{OmegaShift}
\end{equation}
In Fig.~\ref{AccelerationDependence} an example of laminar vortex motions is shown where $[\Omega_\mathrm{c}^*]^2$ is plotted at different accelerations $\dot{\Omega}$, starting from a minimum rate of $10^{-4}$\,rad/s${^2}$. The intercept at $\dot{\Omega} = 0$ gives the true critical value $\Omega_\mathrm{c} = 1.34\,$rad/s, while from the slope $2d/(\alpha R)$ one obtains the mutual friction damping of vortex motion $\alpha = 1.1$ at 1.76\,mK, which is in good agreement with other measurements. This `flight-time' corrected method of identifying $\Omega_\mathrm{c}$ works well for accelerations up to $\dot{\Omega} \sim 0.01\,$rad/s$^2$.

The spiralling vortex motions in the long rotating cylinder after a KH event become most interesting in the turbulent regime below  $0.6\,T_\mathrm{c}$ (see eg. \cite{PNAS}). Here the KH instability is followed by a sudden turbulent burst, where the vortex loops transferred across the AB interface interact in the B phase by reconnecting \cite{Front}. The burst takes place in the vicinity of the AB interface within a space comparable in size to the cylinder radius $R$. It increases the number of spiralling vortices and removes thereby much of the rotating counterflow in this section of the cylinder. The expansion into the vortex-free flow then continues as a spiralling turbulent vortex front with an axial length $\sim R$ \cite{VorFrontHeating,LowT-Front}. It leaves behind a twisted vortex bundle  \cite{TwistedBundle}, a state which later slowly unwinds.

In principle, $\Omega_\mathrm{c}$ can be located by sweeping one of the variables $\Omega$, $T$, $P$ or $I_\mathrm{b}$. In practice, three different techniques have been used. 1) The measurement in Fig.~\ref{KH-Steps} is the most straightforward for locating $\Omega_\mathrm{c}$. Most of the data in this report was measured this way. Here $\Omega$ is increased from zero until the first vortices are detected in B phase, using constant slow acceleration. 2) Since $\Omega_\mathrm{c}(T,P,I_\mathrm{b})$ is highly predictable and often already approximately known, a faster more accurate technique is a rapid increase across $\Omega_\mathrm{c}$ by an increment which can be as small as $\Delta \Omega \sim 0.02\,$rad/s (consult Fig.~\ref{KH-Steps}), followed by a longer waiting time at constant $\Omega$ (exceeding the duration of vortex expansion), to make sure that no further vortices follow. This waiting time can be used for monitoring the NMR line shape, to correct for drifts. 3) Since the heat leak to the sample depends on $\Omega$ \cite{VorFrontHeating}, drifts can be reduced by scanning instead of $\Omega$ the barrier current $I_\mathrm{b}$, \textit{i.e.} by sweeping the current slowly downward until the AB interface becomes unstable in the reduced field gradient.

\section{KH measurements}
\label{Results}
\subsection{General characterization}

\begin{figure}[t]
 \centerline{
  \includegraphics[width=1.1\linewidth]{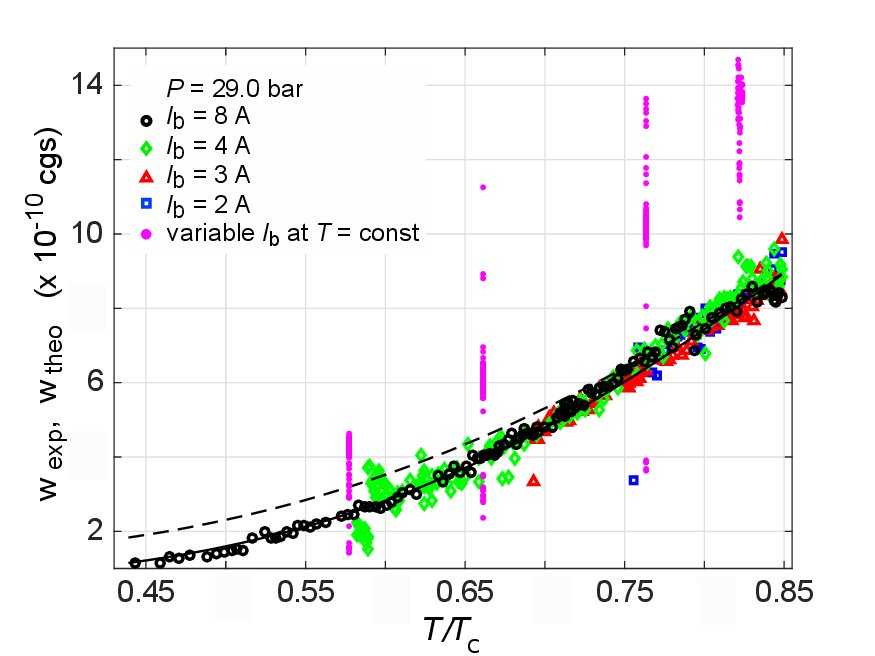}}
\caption{KH measurements at 29.0\,bar pressure. The experimental part $w_\mathrm{exp}$ from Eq.~(\ref{SuperfluidInstability-4}) is plotted as a function of temperature $T/T_\mathrm{c}$. The majority of the data comes from measurements of $\Omega_\mathrm{c}$ as function of temperature at fixed barrier-magnet current $I_\mathrm{b}$. The vertical sets of data points represent measurements at constant temperature as a function of $I_\mathrm{b}$. This data illustrates the total range of the KH phenomenon available with the present experimental setup at different temperatures. The dashed and solid-line curves correspond to  $w_\mathrm{theo}$ calculated without and with the fitted magnetic field dependence of the surface tension, respectively. The comprehensive agreement of the data with the KH model (solid curve) can be conveniently judged with this plot.
 }
\label{KH_w_exp_29bar}
\end{figure}

Fig.~\ref{KH_w_exp_29bar} illustrates measurements of the KH critical velocity $\Omega_\mathrm{c} (T, I_\mathrm{b})$, by plotting $w_\mathrm{exp} (T)$ (from Eq.~(\ref{SuperfluidInstability-4})) as a function of temperature at a fixed pressure of 29\,bar. This plot provides a convenient way of inspecting most of the results. The slowly declining trace of data points represents measurements at different but constant $I_{\mathrm b}$. In contrast, data measured at constant temperature and variable $I_{\mathrm b}$ falls in this plot on vertical lines which have been included to visualize the entire measurable range of the KH instability.

\begin{figure}[tt!]
  \centerline{
  \includegraphics[width=1.1\linewidth]{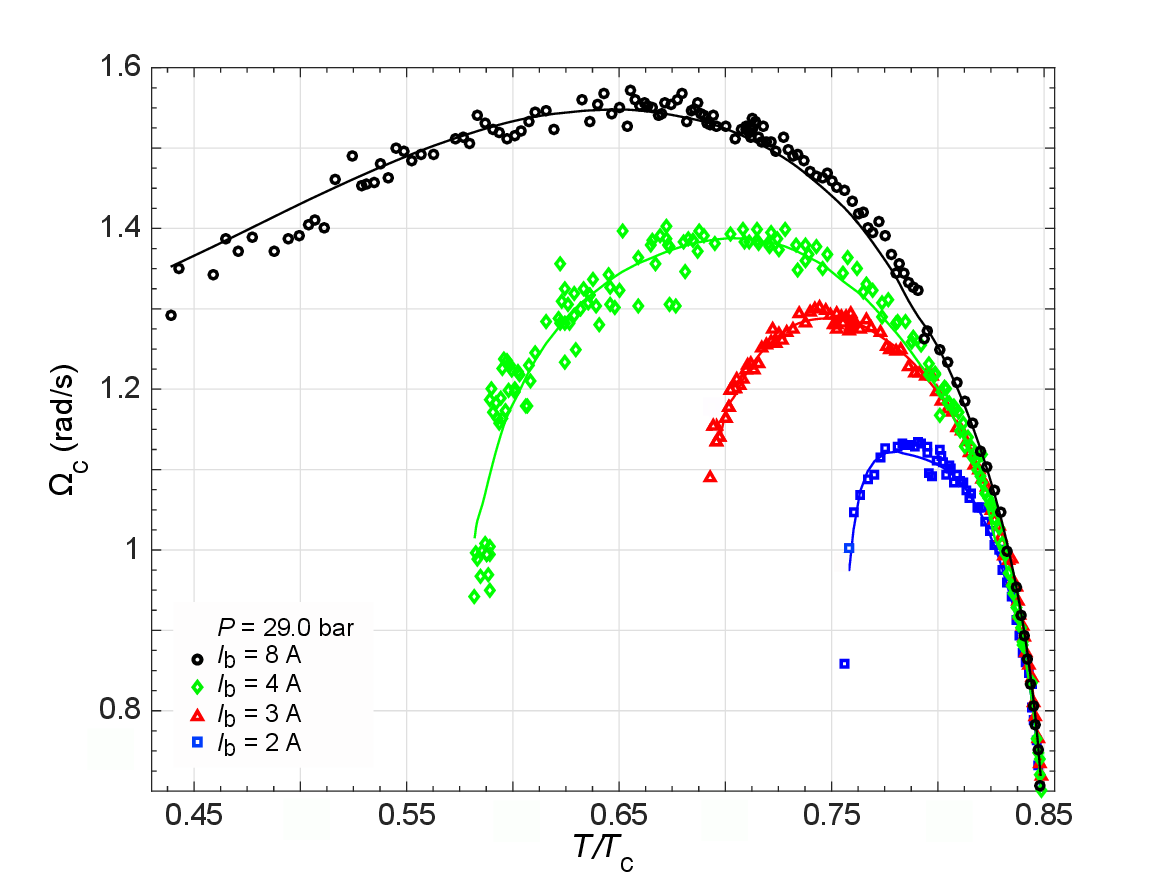}}

\caption{KH critical rotation velocity $\Omega_\mathrm{c} (T)$ at 29\,bar. Here the first KH instability event is plotted, obtained while rotation is slowly increased from zero at constant $T$, $P$, and $I_{\mathrm b}$. Temperature is changed incrementally from one constant value to the next. The different sets of data represent measurements at different fixed currents $I_{\mathrm b}$. The $I_{\mathrm b}$ = (2---4)\,A curves terminate at low temperatures where the AB interface is not maintained at these currents. The solid curves represent fitting (to Eq.~(\ref{SuperfluidInstability-5})) with the magnetic field dependent surface tension.  }
\label{Omega-c-29bar}
\end{figure}

In Fig.~\ref{Omega-c-29bar} a more generic plot is shown with $\Omega_\mathrm{c}$ versus $T$. In this plot the data falls on curves with a  characteristic ``umbrella'' shape, which starts at high pressures ($P > 21.2\,$bar) from $T_\mathrm{AB}$ or at lower pressures from  $T_\mathrm{c}$. Towards low temperatures at large $I_{\mathrm b}$ the curve flattens out towards a constant, while at smaller $I_{\mathrm b}$ the AB interface might not be maintained at low temperatures and the curve plummets down to a termination point.

The ``umbrella'' shape is dictated by the field distribution of the cylindrical barrier solenoid: 1) With decreasing temperature the critical field $H_\mathrm{AB}$ increases approximately parabolically and the AB interface moves towards higher fields inside the solenoid, where it follows the field contour $ H(r,z) = H_\mathrm{AB} (T,P)$. In other words, the location $z_\mathrm{AB}$ of the AB interface moves with decreasing temperature closer to the magnet center and the field gradient increases. Measurements at $I_\mathrm{b} = 8\,$A represent this behavior where the data ultimately at the lowest temperatures tends towards a constant value.  2) At lower $I_\mathrm{b}$, the AB interface reaches the inflection point in the field distribution at some temperature. Below this temperature the gradient $\mid \nabla H \mid_{H = H_\mathrm{AB}}$ starts decreasing, until at some low temperature limit the A phase disappears. This behavior is exemplified by the data measured at (2---4)\,A in Fig.~\ref{Omega-c-29bar}.

In the $w_\mathrm{exp} (T)$ plot in Fig.~\ref{KH_w_exp_29bar} all data collapses on one common curve which at high pressures is monotonically decreasing with decreasing temperature.  At low pressures, where the zero-field $T_\mathrm{AB}$ moves to $T_\mathrm{c}$, the curve displays an initial steep rise just below $T_\mathrm{c}$, where the rapid increase in $\Omega_\mathrm{c} (T)$ first dominates, before settling down on a slowly descending dependence with decreasing temperature.

The vertical columns of data points in Fig.~\ref{KH_w_exp_29bar} represent measurements at constant temperature (and thus fixed $H_\mathrm{AB} $), while the current $I_{\mathrm b}$ is varied. They display the range of variation in $w_\mathrm{exp}$ obtainable with the present barrier solenoid where the maximum operating current in the CuNi-clad filamentary NbTi superconducting wire is limited to 8\,A. This limits the maximum of the vertical column, which corresponds to the situation when $z_\mathrm{AB}$ is furthest away from the magnet center (well beyond the inflection point of the field distribution $H_z (z)$). At lower $I_{\mathrm b}$, $z_\mathrm{AB}$ resides closer to the  magnet center. The bottom end of the column corresponds to the lowest current $I_{\mathrm b}$ at which the magnetic restoring force $\propto H_\mathrm{AB} (T) \, \mid ~\!\nabla H~\mid_{z = z_\mathrm{AB}}$ still maintains an interface and $z_\mathrm{AB}$ is closest to the magnet center. Owing to the fixed upper limit $I_{\mathrm b} \leq 8\,$A and the termination at low $I_{\mathrm b}$, which moves to higher $I_{\mathrm b}$ values when the temperature is reduced, the height of the vertical data columns decreases towards low temperatures.

Hence both types of measurement series terminate in a collapse of the interface if the magnetic restoring force becomes too weak. Careful measurements around such termination points display more complex behavior than described above \cite{LT23-FieldConf}. When the termination is approached from above, the thickness of the A-phase layer in Fig.~\ref{Setup} decreases until a hole is punched into the final thin A-phase membrane. In this final state, the A-phase volume collapses to a narrow annulus coating the cylinder wall, where the interface is still sustained by a slightly larger restoring force. Owing to this change in the topology of the AB interface and the associated barriers in nucleating the first order AB phase transition, $\Omega_\mathrm{c} (T, I_{\mathrm b})$  displays in the vicinity of a termination point both thermal and magnetic hysteresis. Thus around a termination point the measured $\Omega_\mathrm{c}$  appears to show more scatter and lie below the regular dependence (see \textit{e.g.} the $I_{\mathrm b} = 4\,$A data at low temperatures in Fig.~\ref{KH_w_exp_29bar}).

\begin{figure}[t!]
  \centerline{
  \includegraphics[width=1.1\linewidth]{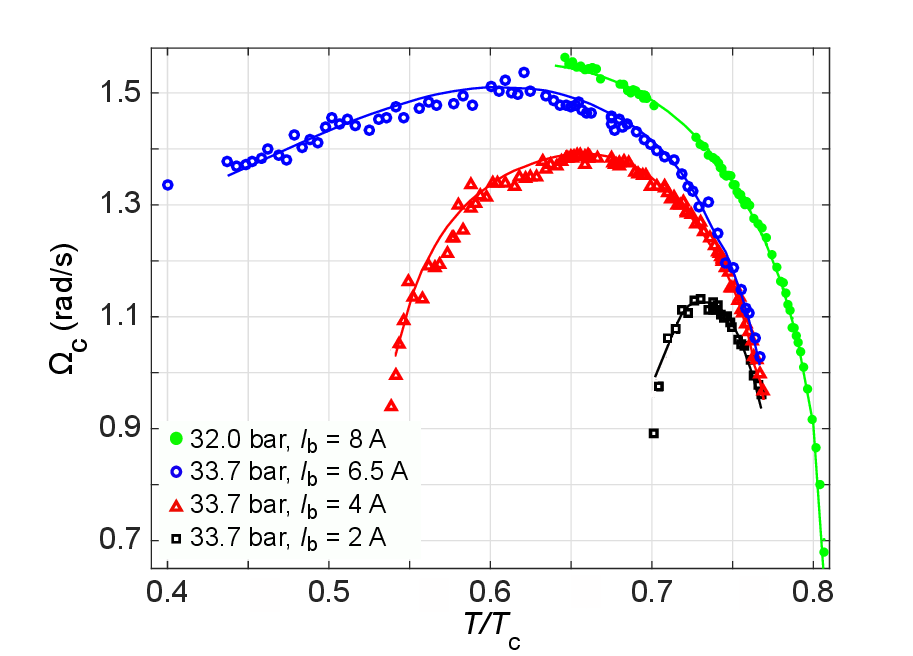}}

\caption{Critical rotation velocity of KH instability at 32.0 and 33.7\,bar. The same conventions are used as in Fig.~\ref{Omega-c-29bar}. The 32.0\,bar measurements were conducted after the solid $^3$He plug in the fill line of the sample container had accidentally slipped. The slip reduced the pressure to an unknown value. The assignment $(32.0 \pm 0.25)\,$bar was determined from the measured temperature of the B $\rightarrow$ A transition at zero field: $T_\mathrm{AB}/T_\mathrm{c} \simeq 0.81$.
}
\label{Omega-c-33.7bar}
\end{figure}

The high pressure data in Figs.~\ref{Omega-c-29bar} and \ref{Omega-c-33.7bar} was collected in the AB configuration of phase stacking (see Fig.~\ref{Setup}), where the instability of only one AB interface can be recorded. The measurements at 10.2\,bar (Fig.~\ref{Omega-c-10.2bar}) and at zero pressure (Fig.~\ref{Omega-c-0bar}) were performed in the BAB stacking configuration, where the independently occurring instabilities at the two AB interfaces were separately monitored. When comparing a measured data point of $\Omega_\mathrm{c}$ to its calculated estimate [obtained from Eq.~(\ref{SuperfluidInstability-5})], the input data are $T$, $I_{\mathrm b}$, and $P$. Since the experimental setup is not exactly identical with respect to the two AB interfaces (see Fig.~\ref{BarrierField}), $\Omega_\mathrm{c}$ is calculated separately for the upper and lower phase boundaries. In Fig.~\ref{Omega-c-10.2bar} both data sets have been plotted as solid-line curves. The agreement of the two curves at each value of $I_{\mathrm b}$ is good which serves to show that the magnetic field configuration is in good control.

Below we analyze the two types of measurements, examining first  measurements at constant $I_{\mathrm b}$ and comparing the result then to measurements at constant $T$.

\subsection{KH instability at constant magnet current}
\label{const_Ib}

\begin{figure}[t]
  \centerline{
  \includegraphics[width=1.1\linewidth]{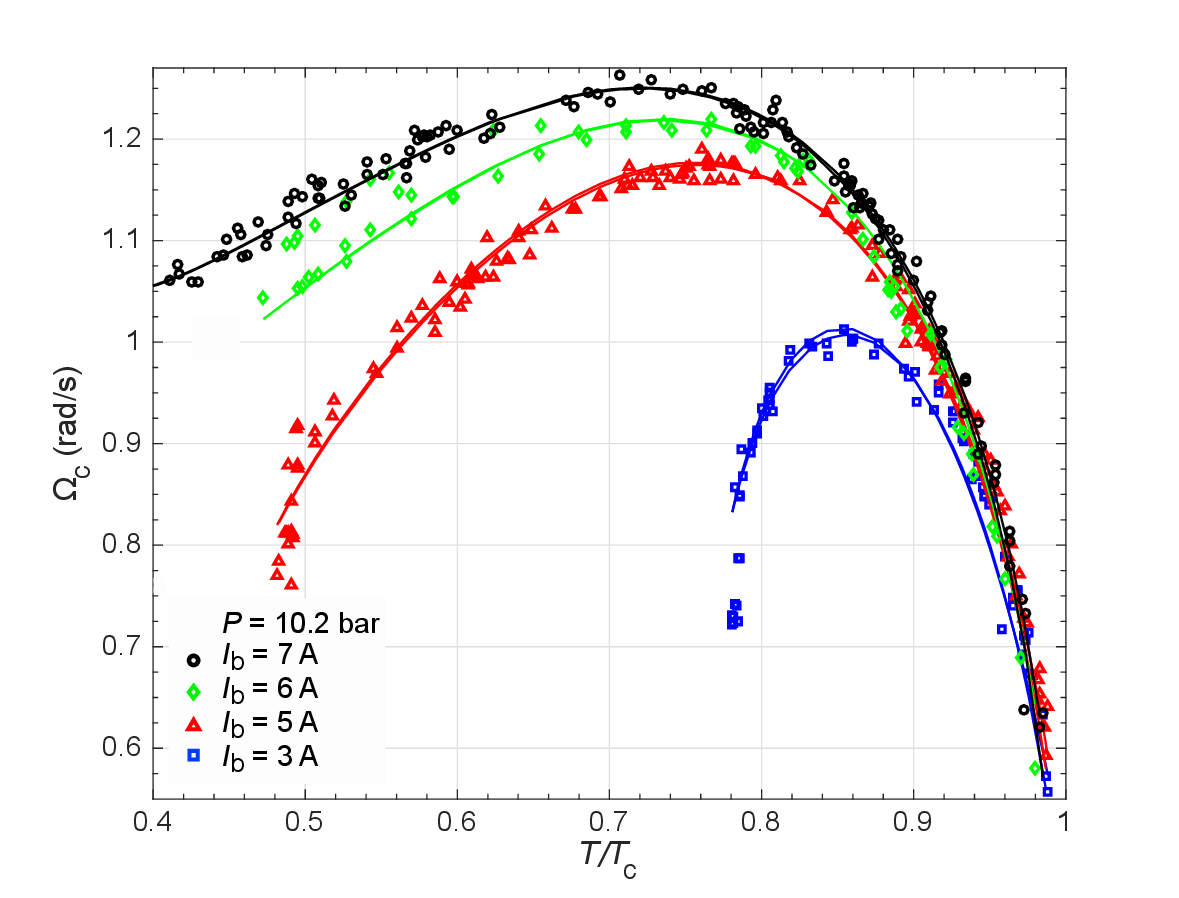}}

\caption{KH rotation velocity at 10.2\,bar, measured in the BAB phase stacking configuration. No systematic differences were noted between $\Omega_\mathrm{c}$ values recorded at the two interfaces. For clarity, the figure does not identify from which section a data point originates. $\Omega_\mathrm{c}$ is calculated for both interfaces separately (from Eq.~(\ref{SuperfluidInstability-5})) which gives two closely agreeing values plotted as solid-line curves for each value of $I_{\mathrm b}$. }
\label{Omega-c-10.2bar}
\end{figure}

\begin{figure}[t]
  \centerline{
  \includegraphics[width=1.1\linewidth]{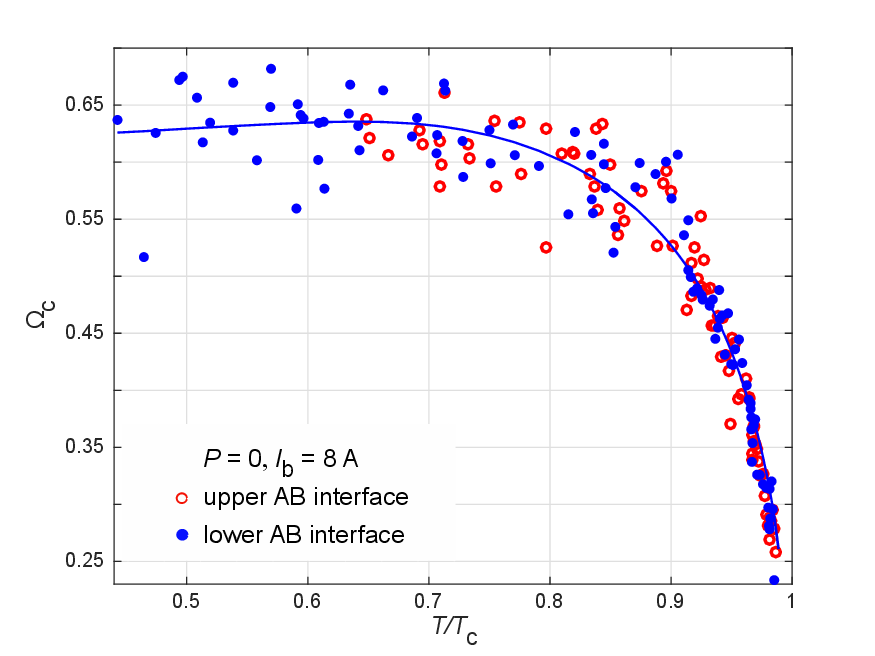}}

\caption{KH rotation velocity at zero pressure. The instabilities of the two AB interfaces (see Fig.~\ref{Setup}) occur independently and their critical velocities show no systematic deviations from a common dependence. The scatter is larger than at elevated pressures, owing to lower rotation velocities and smaller frequency shifts in the NMR measurement \cite{Textures}. }
\label{Omega-c-0bar}
\end{figure}

Here we describe how the solid-line curves in Figs.~\ref{KH_w_exp_29bar}---\ref{Omega-c-0bar} have been generated, \textit{i.e.} our fitting procedure for comparing the measured data to the instability criterion (\ref{SuperfluidInstability-2}). At constant pressure, the measured $\Omega_\mathrm{c} (T,P,I_\mathrm{b})$ is compared to its calculated estimate from
\begin{equation}
\Omega_\mathrm{c} (T) = {1 \over R_\mathrm{eff}} \; \left[ {4 \, H_\mathrm{AB} \; \mid \! \nabla H \! \mid_{H = H_\mathrm{AB}} \frac {\sigma \; \Delta \chi} {\rho_\mathrm{s,eff}^2} } \right] ^{1/4} \;.
\label{SuperfluidInstability-5}
\end{equation}
As noted in Sec.~\ref{MagFieldDep}, reasonable agreement can be reached at high temperatures, but towards low temperatures the extrapolation of Eq.~(\ref{SuperfluidInstability-5}) increasingly overestimates $\Omega_\mathrm{c}$, being above the measurements by about 10\,\% at $0.5\,T_\mathrm{c}$ at the pressures 29 -- 34\,bar (see dashed curve in Fig.~\ref{KH_w_exp_29bar}). We attribute the deviation to the magnetic polarization of B phase at high $H_\mathrm{AB} (T)$ fields, with the largest contribution resulting from the surface tension $\sigma_\mathrm{AB} (T,P,H)$.

The critical field $H_\mathrm{AB} (T,P)$ is taken from Ref.~\cite{Hahn}, the field gradient $ \nabla H \! \mid_{H = H_\mathrm{AB}}$ is calculated from the information given for the barrier solenoid in Ref.~\cite{LT23-FieldConf}, the B-phase susceptibility $\chi_\mathrm{B \perp} (T,P)$ is interpolated from the results measured by Scholz \cite{Scholz}, and the superfluid densities $\rho_{\parallel}$ and $\rho_{\perp}$ are calculated numerically in the weak coupling approximation. The magnetic field dependences are thus approximately accounted for, except in the surface tension, which is the quantity we determine by fitting.

The low-field surface tension is introduced in the form $\sigma_\mathrm{AB} (T,P) = \sigma_0 (P) (1-T/T_\mathrm{c})^{3/2}$. The initial estimate of $\sigma_0 (P)$ is obtained by fitting the high temperature data at $H_\mathrm{AB} \leq 0.1\,$T for each constant pressure measurement separately. The residual deviation beyond this estimate proves to be $\propto H_\mathrm{AB}^2$ at lower temperatures; in other words a good fit at low temperatures is obtained assuming a surface tension of the form $\sigma (T,P,H_\mathrm{AB}) = \sigma_0 (P) \; (1-T/T_\mathrm{c})^{3/2} \; (1-a H_\mathrm{AB}^2)$. Since the experimental data sets at each pressure are sizeable, the final step is to improve the fitting for $\sigma_0 (P)$ and $a$ by searching for an overall minimum squared deviation using a smooth polynomial for $\sigma_0 (P)$ as a function of pressure. The error analysis shows that the dominant uncertainty arises from locating properly the best combination of these two parameter values in the shallow minimum of the sum of the squared deviations. This procedure leads to uncertainty limits of a few percent for $\sigma_0 (P)$ and somewhat larger for $a$, $\lesssim 10\,$\%.

\begin{figure}[t]
  \centerline{
  \includegraphics[width=0.9\linewidth]{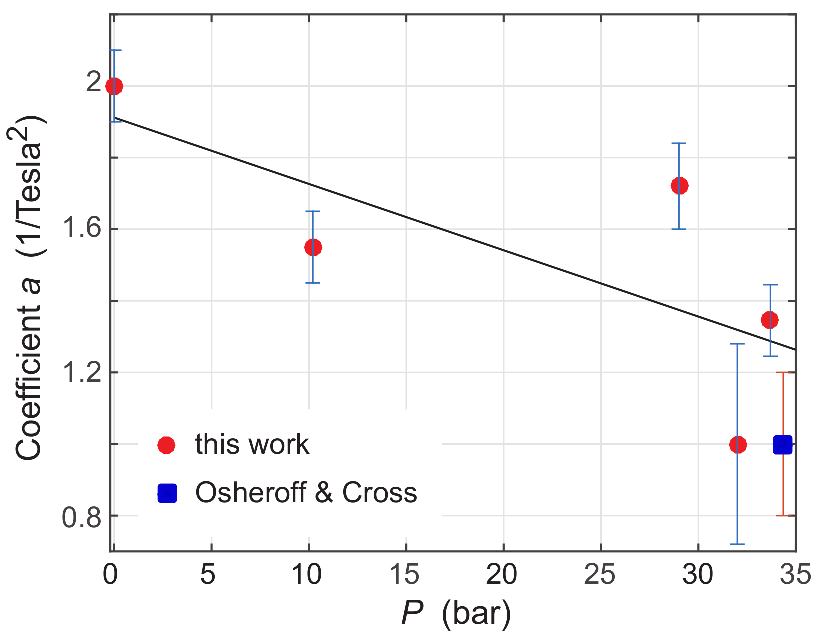}}

\caption{ Surface tension of AB interface in magnetic field. The coefficient $a$ in the parabolic suppression of the surface tension with magnetic field is plotted at different sample pressures: $\sigma_\mathrm{AB} (T,P,H) = \sigma_0 (P) \; (1-T/T_\mathrm{c})^{3/2} \; (1-a H^2)$. The fitted line corresponds to $ a =1.91 - 0.0185 P$/bar (in T$^{-2}$). The square data point
represents a fit to the measurements in Ref.~\cite{Osheroff} (see Fig.\,\ref{OsheroffResult}). The error bars reflect the overall uncertainty in fitting the measurements at each pressure. }
\label{FieldDependence}
\end{figure}

While searching for the best value of $\sigma_0 (P)$, it is actually the scale factor $\Omega_\mathrm{c} \propto \sigma_0^{1/4}\, / \, R_\mathrm{eff}$ which is fitted. The only reported low temperature measurement of surface tension \cite{Bartkowiak} gives $\sigma(0.15 \, T_\mathrm{c}, P = 0) = (3.03 \pm 0.28) \cdot 10^{-9}\,$J/m$^2$ in a field of $H_\mathrm{AB} = 338\,$mT. We use this value to extract from the fitted scale factor at zero pressure an effective radius of $R_\mathrm{eff} = 2.67\,$mm. This number is in line with $R_\mathrm{eff} = 2.5\,$mm determined in Fig.~\ref{KH-Steps} or with $R_\mathrm{eff} = 2.6\,$mm measured in Ref.~\cite{InterfaceVortexSheet}. With the assumption that $R_\mathrm{eff}$ is pressure independent, we extract from the fitted scale factors the surface tension $ \sigma_0 (P) \simeq (5.02 + 3.40 P - 0.112 P^2 + 0.00123 P^3)\, 10^{-9}$J/m$^2$ ($P$ in bar). We remind that this expression represents the surface tension measured at low field, $H < 0.1$\,T, and not a true zero-field-limit value. In fact, at pressures below the polycritical point the AB interface is not even stable at $H=0$. The first-order magnetic field dependent correction is depicted in Fig.~\ref{FieldDependence}


\begin{figure}[t]
  \centerline{
  \includegraphics[width=1\linewidth]{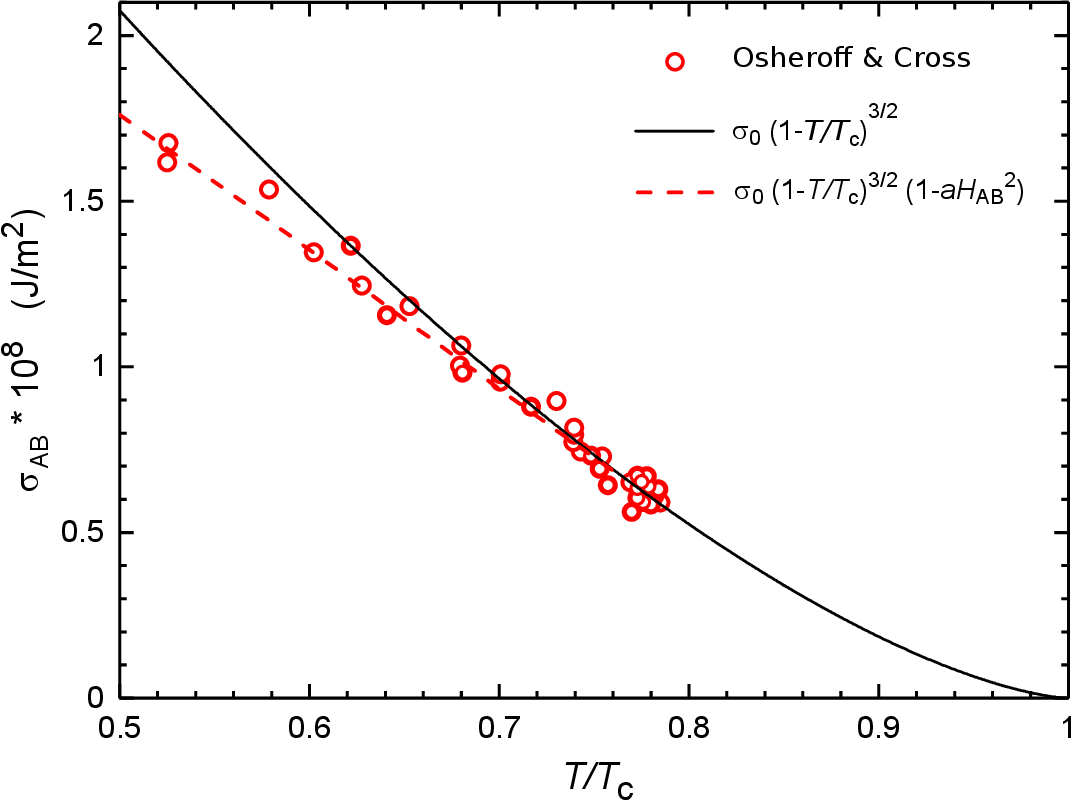}}

\caption{Measurements of AB interface surface tension at melting pressure (34.4 bar) in Ref.~\cite{Osheroff}. The solid curve represents $\sigma_\mathrm{AB} (T,H=0) = \sigma_0 \; (1-T/T_\mathrm{c})^{3/2}$, derived from a fit to those data points with $T > 0.76\,T_\mathrm{c}$. This gives $\sigma_0 = 5.87*10^{-8}$\,J/m$^2$. The dashed curve for $\sigma_\mathrm{AB} (T,H)$ is obtained by fitting for the magnetic field correction using all data points. This gives the data point marked with a square in Fig.~\ref{FieldDependence}.}
\label{OsheroffResult}
\end{figure}

Adjusting our analysis to the surface tension measured in Lancaster thus appears to give consistent answers. In Ref.~\cite{Osheroff} Osheroff and Cross measure the surface tension at melting pressure from $T_\mathrm{AB} = 0.78\,T_\mathrm{c}$ down to $0.53\,T_\mathrm{c}$. Their data points are plotted in Fig.~\ref{OsheroffResult}. A fit to their high-temperature points with $T > 0.76\,T_\mathrm{c}$ yields $\sigma (T) = 5.87 \cdot 10^{-8} \; (1-T/T_\mathrm{c})^{3/2}\,$J/m$^2$, while from our fitted polynomial we get the prefactor $3.95 \cdot 10^{-8} \,$J/m$^2$, a 30\,\% smaller value. The reason for this difference has not been identified. In contrast, our parabolic magnetic field correction with $a \simeq 1$T$^{-2}$ can be regarded to be consistent with Osheroff's data, as seen in Figs.~\ref{FieldDependence} and \ref{OsheroffResult}.

In Ref.~\cite{Halperin} the consistency of the Osheroff - Cross surface tension with Thuneberg's GL expansion is examined, by searching for a proper combination of strong-coupling corrected GL $\beta$ parameters. Such curves have the temperature dependence of the black curve in Fig.~\ref{OsheroffResult}. However, temperatures as low as $\sim 0.5 \, T_\mathrm{c}$ in Ref.~\cite{Osheroff} are not adequately represented by a zero-field GL expansion. This is seen in Fig.~\ref{OsheroffResult}, where the magnetic polarization from $H_\mathrm{AB} (T)$ enforces an increasing wedge between the black and red curves. We recommend that the low temperature data points below $\sim 0.65\,T_\mathrm{c}$ should be excluded from the fitting in Ref.~\cite{Halperin}, to improve the reliability of the fitting procedure. By comparing Fig.~\ref{OsheroffResult} to Fig. 4 in Ref.~\cite{Halperin}, one can estimate how this affects the choice of the $\beta$ parameters.

Finally we note in passing that the A-phase section was generally prepared to contain the equilibrium number of doubly quantized vortex lines which have a continuous order parameter distribution  \cite{DoubleQuantumVortex}. For reference, in some cases also a second continuous vortex texture, the equilibrium vortex sheet, was grown \cite{VortexSheet}. Within the experimental scatter $\Omega_\mathrm{c} (T)$ was found to be unaffected by the choice of the A-phase vortex structure. This conclusion is expected: the instability in Eq.~(\ref{SuperfluidInstability-2}) depends on the tangential flow velocities at the interface which are not affected by the vortex structure if the flow conditions remain otherwise unchanged.

\begin{figure}[t]
  \centerline{
  \includegraphics[width=0.95\linewidth]{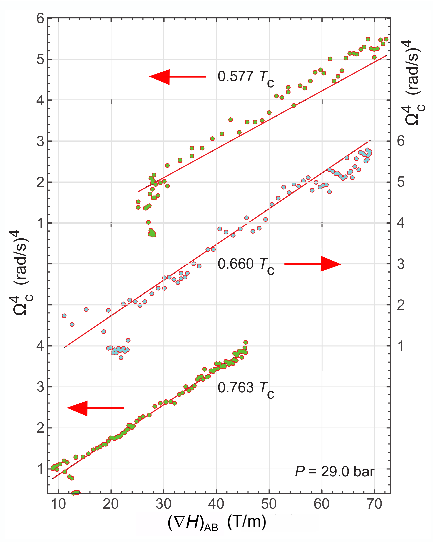}}

\caption{KH instability measured at constant temperature as a function of the current $I_{\mathrm b}$ in the barrier magnet at 29.0~bar. At all three temperatures the low-current termination includes data points where the A phase volume is not a singly connected complete layer. The lines represent Eq.~(\ref{SuperfluidInstability-5}). (Note that the vertical scales of the two upper data sets have been moved upward).}
\label{Om_gradH29barComposite}
\end{figure}

\subsection{KH instability at constant temperature}
\label{const_T}

\begin{figure}[t]
  \centerline{
  \includegraphics[width=0.95\linewidth]{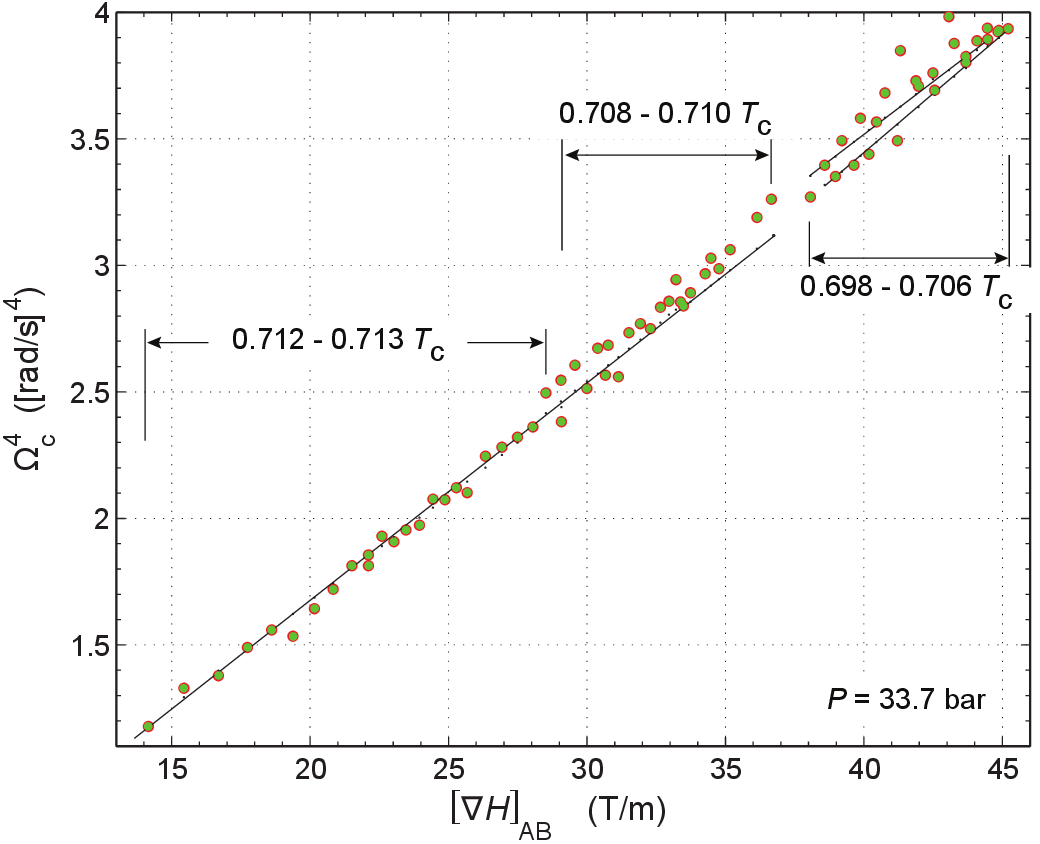}}

\caption{KH instability measured at a constant temperature $\sim 0.70\,T_\mathrm{c}$ at 33.7 bar. The data was collected in three separate measuring sessions within slightly differing temperature ranges, as indicated in the plot. The lines with slightly differing slopes display Eq.~(\ref{SuperfluidInstability-5}).
}
\label{Om_gradH33.7bar1.77mK}
\end{figure}

In measurements at constant temperature as a function of $I_{\mathrm b}$, the linear dependence $\Omega_\mathrm{c}^4 \propto \; \mid\!\nabla H \! \mid_{H = H_\mathrm{AB}}$ in Eq.~(\ref{SuperfluidInstability-5}) can be tested. This linear relation is illustrated in Fig.~\ref{Om_gradH29barComposite} at three different temperatures. At each of these temperatures the field $H_\mathrm{AB}$ remains constant, while $I_{\mathrm b}$ increases and the location of the AB interface moves further away from the magnet center.  The measurement of the slope requires accurate temperature control, as demonstrated by the illustration in Fig.~\ref{Om_gradH33.7bar1.77mK}. The instability can be traversed by sweeping $\Omega$ upward at constant $I_{\mathrm b}$ or by sweeping $I_{\mathrm b}$ downward at constant $\Omega$. In both cases rotation has to be stopped after each instability event, to remove the vortices from the B phase sections and to initialize the measuring setup for the next round of measurements. As the heat leak depends on $\Omega$ \cite{Hosio-PRB}, maintaining precise temperature stability requires careful work. In Fig.~\ref{Om_gradH33.7bar1.77mK} a 2\,\% variation in the measuring temperature is seen to lead to a 5\,\% uncertainty in the determination of the slope $\Omega_\mathrm{c}^4$ versus $\nabla H $.

A further concern in the measurements of Fig.~\ref{Om_gradH33.7bar1.77mK} is magnetic remanence trapped in superconducting materials, which might contribute to the value of the field gradient. The most sensitive place to search for remanent flux motion is the vicinity of a low-current termination point (see Ref.~\cite{LT23-FieldConf}). However, sweeping $I_{\mathrm b}$ up and/or down does not increase the scatter of the measured $\Omega_\mathrm{c}$ in Fig.~\ref{Om_gradH33.7bar1.77mK}, which means that any trapped flux has to be strongly pinned. Nevertheless, it is possible that persistent remanent fields are trapped in the initial cool down of the apparatus while charging the various superconducting magnets. These could explain possible systematic deviations between different cool-downs. However in this context it suffices to summarize that we find reasonable agreement of the data as a function of $I_{\mathrm b}$ with Eq.~(\ref{SuperfluidInstability-5}) using the surface tension obtained in Sec.~\ref{const_Ib}.

\section{Conclusion}
\label{Analysis}

The KH instability of the AB interface takes place between two stable bulk states of the superfluid $^3$He order parameter manifold. The interface resides at the magnetic field $H_\mathrm{AB} (T,P)$ and is firmly localized by a steep field gradient $\mid \! \nabla H \! \mid_{H = H_\mathrm{AB}}$. At high temperatures, corresponding to $H_\mathrm{AB} < 0.1$\,T, the theoretical instability criterion (\ref{SuperfluidInstability-2}) provides a good explanation of the measurements, with superfluid $^3$He properties expected for zero magnetic field. At low temperatures fields as large as $H \lesssim 0.6$\,T are encountered and an increasing influence of the magnetic polarization of the B-phase properties becomes apparent as a sizeable reduction of the interface stability.

Little quantitative experimental information exists for comparing the magnetic deformation of the energy gap and its influence on the properties responsible for the interface stability. The first attempt to improve agreement by considering the boundary conditions at the interface and changes in the B-phase order parameter texture as a function of the applied field showed \cite{AnisotropicKH} that the stability criterion had to be amended by a renormalized superfluid density ~(\ref{EffectiveDensity}). However, comparison with measurements also revealed that the magnetic-field-induced reduction in the surface tension becomes an even more important consideration. Its first-order parabolic correction with magnetic field has here been extracted at different pressures down to a temperature of $0.4\,T_\mathrm{c}$. It is found to be in good agreement with the value of surface tension measured by Bartkowiak \textit{et al.} \cite{Bartkowiak} at zero pressure and $0.15\,T_\mathrm{c}$.

This agreement lends support to our conclusion that the KH instability provides a new noninvasive technique for extracting the AB-interface surface tension, a technique which is different from the classic measurement of the interface popping through an oriﬁce, the case in Refs.~\cite{Osheroff,Bartkowiak}. Thus this work becomes the first measurement where the AB-interface surface tension is comprehensively determined at different pressures over most of the temperature regime. The magnetic field dependent correction of the AB interface energy in Fig.~\ref{FieldDependence} can now be used for simple estimates of the reduction in B-phase stability in increasing field. But more importantly, the surface tension is calculable at low temperatures and the present results, we hope, provide an incentive for such comparison.

\textit{\textbf{Acknowledgements}}:--We are indebted to many generations of graduate students who contributed to this study. We thank our late colleague Nikolai Kopnin \cite{Kopnin} for encouragement, similarly Erkki Thuneberg \cite{Thuneberg_sigma} and Grigori Volovik \cite{Volovik} for their interest. Financial support was provided by the Aalto University School of Science and its research infrastructure Low Temperature Laboratory.

\end{document}